\documentclass{aa}

\usepackage{graphicx}
\usepackage{txfonts}
\usepackage{upgreek}
\usepackage[colorlinks,citecolor=blue,urlcolor=blue,filecolor=blue,linkcolor=blue]{hyperref}

\makeatletter
\renewcommand*\aa@pageof{, page \thepage{} of \pageref*{LastPage}}
\makeatother

\defcitealias{itu_p452_16}{Rec.~ITU-R~P.452-16}
\defcitealias{itu_p452_17}{Rec.~ITU-R~P.452-17}
\defcitealias{itu_ra2259_1}{Rep.~ITU-R~RA.2259-1}
\defcitealias{itu_m1583_1}{Rec.~ITU-R~M.1583-1}
\defcitealias{itu_p1144_9}{Rec.~ITU-R~P.1144-9}
\defcitealias{itu_ra769_2}{Rec.~ITU-R~RA.769-2}
\defcitealias{itu_ra1513_2}{Rec.~ITU-R~RA.1513-2}
\defcitealias{itu_ra1631_0}{Rec.~ITU-R~RA.1631-0}
\defcitealias{itu_f1766_0}{Rec.~ITU-R~F.1766-0}
\defcitealias{itu_s1586_1}{Rec.~ITU-R~S.1586-1}
\defcitealias{itu_ra2332}{Rep.~ITU-R~RA.2332}
\defcitealias{cispr11}{EN\,550011 (CISPR-11)}
\defcitealias{ecc_report_171}{ECC~Report~171}
\defcitealias{ecc_report_247}{ECC~Report~247}
\defcitealias{ecc_report_271}{ECC~Report~271}
\defcitealias{ECSS-E-ST-20-07C}{ECSS-E-ST-20-07C}
\defcitealias{MSFC-SPEC-521C}{MSFC-SPEC-521C}
\defcitealias{radioregs}{Radio Regulations}

\usepackage{xcolor}
\usepackage{ulem}

\begin{document}
\title{Unintended electromagnetic radiation from Starlink satellites detected with LOFAR between 110 and 188~MHz}
\titlerunning{Unintended electromagnetic radiation from Starlink satellites detected with LOFAR}

\author{
  F.\,Di\,Vruno\inst{\ref{skao},\ref{craf}}\fnmsep\thanks{Member of the IAU Centre for the Protection of the Dark and Quiet Sky from Satellite Constellation Interference (IAU CPS).}
  \and
  B.\,Winkel\inst{\ref{mpifr},\ref{craf}}\fnmsep$^\star$
  \and
  C.\,G.\,Bassa\inst{\ref{astron}}\fnmsep$^\star$
  \and  
  G.\,I.\,G.\,J\'{o}zsa\inst{\ref{mpifr},\ref{craf},\ref{rhodes}}\fnmsep$^\star$
  \and  
  M.\,A.\,Brentjens\inst{\ref{astron}}
  \and  
  A. Jessner\inst{\ref{mpifr}}
  \and  
  S. Garrington\inst{\ref{jbo}}\fnmsep$^\star$
}

\institute{
  Square Kilometre Array Observatory, Lower Withington, Macclesfield, Cheshire, SK11 9FT, United Kingdom\\ \email{Federico.DiVruno@skao.int}\label{skao}
  \and
  European Science Foundation, Committee on Radio Astronomy Frequencies, 1, quai Lezay Marnésia BP 90015, F-67080 Strasbourg Cedex, France\label{craf}
  \and
  Max-Planck-Institut f\"ur Radioastronomie, Auf dem H\"ugel 69, 53121
  Bonn, Germany\label{mpifr}
  \and
  ASTRON, Netherlands Institute for Radio Astronomy, Oude
  Hoogeveensedijk 4, 7991 PD Dwingeloo, The Netherlands
  \label{astron}
  \and
  Department of Physics and Electronics, Rhodes University, PO Box 94, Makhanda, 6140, South Africa\label{rhodes}
  \and
  Jodrell Bank Centre for Astrophysics, Department of Physics and Astronomy, University of Manchester, Manchester M13 9PL, United Kingdom
  \label{jbo}
}

\date{Received March 10, 2023; accepted May 12, 2023}

\abstract{We report on observations of 68 satellites belonging to the SpaceX Starlink constellation with the LOFAR radio telescope. Radiation associated with Starlink satellites was detected at observing frequencies between 110 and 188~MHz, which is well below the 10.7 to 12.7~GHz radio frequencies used for the downlink communication signals. A combination of broad-band features, covering the entire observed bandwidth, as well as narrow-band (bandwidth\,$<12.2$~kHz) emission at frequencies of 125, 135, 143.05, 150, and 175~MHz, was observed. The presence and properties of both the narrow- and broad-band features vary between satellites at different orbital altitudes, indicating possible differences between the operational state of, or the hardware used in, these satellites. While the narrow-band detections at 143.05~MHz can be attributed to reflections of radar signals from the French GRAVES Space Surveillance Radar, the signal properties of the broad- and narrow-band features at the other frequencies suggest that this radiation is intrinsic to the Starlink satellites and it is seen for 47 out of the 68 Starlink satellites that were observed. We observed spectral power flux densities vary from 0.1 to 10~Jy for broad-band radiation, to 10 to 500~Jy for some of the narrow-band radiation, equivalent to electric field strengths of up to $49~\mathrm{dB}\left[\mu\mathrm{V}\,\mathrm{m}^{-1}\right]$ (as measured at a 10~m distance from the satellites, with a measurement bandwidth of 120~kHz). In addition, we present equivalent power flux density simulations of the full Starlink phase 1 constellation, as well as other satellite constellations, for one frequency band allocated to radio astronomy by the International Telecommunication Union (ITU). With these, we calculate the maximum radiation level that each satellite constellation would need to have to comply with regulatory limits for intended emissions in that band. However, these limits do not apply if the radiation is unintended, that is to say if it does not originate from intentionally radiated signals for radio communication or other purposes. We discuss the results in light of the (absence of) regulations covering these types of unintended electromagnetic radiation and the possible consequences for astronomical radio observations.}

\maketitle

\section{Introduction}

Modern radio astronomy has profited greatly from advances in technology. Astronomical radio receivers nowadays are often operated with large fractional bandwidths (bandwidth $\Delta\nu$ over observing frequency $\nu$ in excess of $\Delta\nu/\nu>50$\%; e.g.\ \citealt{torne17,hobbs+20}), increased sensitivity, and aperture (e.g.\ \citealt{jonas+16}), as well as wider fields of view (e.g.\ \citealt{johnston+07,hwg+13}). At the same time, the numerical capabilities of digital back ends have enormously increased owing to field programmable gate arrays (FPGAs) or graphics processing units (GPUs) that allow one to implement special-purpose algorithms in flexible hardware boosting the processing speeds. This allows one to record data with unprecedented temporal and spectral resolution, which benefits spectroscopy, pulsar, and very large baseline interferometry (VLBI) observations alike.

However, astronomy is not alone in utilising the radio spectrum. There is a huge number of applications, such as radio and TV broadcasts, high-speed wireless communications (e.g.\ cell phone networks and WiFi), or radars, which require access to the spectrum. Any type of radio communication and intended radio transmissions is regulated to avoid a situation where different operators -- when using the same or nearby frequencies -- create interference on each other’s systems. This regulation of the radio spectrum is handled at the national level by national radio administrations; however, as radio waves do not care for national borders, international rules are required for harmonisation. The Radiocommunication sector of the International Telecommunication Union (ITU-R) is the top level organisation that takes care of this international regulation. It is a specialised agency of the United Nations. The ITU-R publishes the \citetalias{radioregs} (RR), which is an international treaty and member states are expected to transform the RR into national law.

The ITU-R recognised radio astronomy as a service -- the radio astronomy service (RAS) -- already in 1959 and allocated bands in the radio spectrum to it. Unfortunately, the bands that are allocated to the RAS are relatively sparse and narrow -- for spectral-line observations, the majority of the reserved bands only cover the typical Milky Way Doppler shifts. Also, the total amount of spectrum that is allocated to the RAS is not considered to be sufficient for modern radio astronomical research by most scientists. Below 4~GHz, only 5\% of the radio spectrum is allocated to radio astronomy at various levels of protection. If only primary allocations (the highest level of protection) are considered, as little as 1.6\% is allocated to the RAS. For more details about the regulatory process, the radio astronomy service and its protection, we refer readers to the ITU Handbook on Radio Astronomy \citep{itu_handbook}, the CRAF Handbook for Radio Astronomy \citep{craf05}, and the Handbook of Frequency Allocations and Spectrum Protection for Scientific Uses \citep{corf_handbook}.

Not all radiation produced from electronic devices is subject to ITU-R regulations. To a large extent, the RR only cover the so-called emissions, which refer to the radiation that is directly related to the intentional use of radio frequencies in a system (for the purpose of communications, remote sensing, radionavigation, etc). This obviously includes the wanted signals, but also unwanted emission: spectral sidelobes including harmonics and intermodulation products that are an inevitable by-product of the generation of the wanted transmission. Unwanted emission is a consequence of the signal amplification or mixing, the chosen modulation scheme, etc. Both the wanted and unwanted emissions are regulated in the RR. But there is yet another source of electromagnetic radiation present in any electrical device (or system), which is related at its most fundamental level to the acceleration and deceleration of charges in any electrical or electronic circuits and not necessarily related to the generation of wanted radio signals. As the RR did not coin a regulatory term for this, hereafter we refer to this as unintended electromagnetic radiation (UEMR); it is worth mentioning that in engineering this radiation can be referred to as electromagnetic interference (EMI). UEMR can appear, for example, as the product of current loops in switching mode power supplies, communication signals in unbalanced or mismatched transmission lines, fast switching signals in printed circuits, and actuating electromechanical circuits, etc. Basically any electrical circuit generates some level of UEMR.

UEMR is not explicitly regulated at the ITU-R level, though other standardisation organisations have filled the gap. The Comit\'{e} International Sp\'{e}cial des Perturbations Radio\'{e}lectriques (CISPR\footnote{English: International Special Committee on Radio Interference}), which is a part of the International Electrotechnical Commission (IEC), sets standards for all kinds of terrestrial electrical and electronic devices in order to control electromagnetic interference. Unlike the RR, CISPR standards refer not only to radiocommunication systems but all kinds of electronic devices. Furthermore, the standards also cover measurement procedures, which are used to determine the level of UEMR produced by a device under test.

Unlike intended radio emission, UEMR is not clearly specified by a centre frequency, output power and bandwidth, yet it has some characteristics worth mentioning: i) its radiated power is normally several orders of magnitude lower than any intentional radiation; ii) UEMR is usually not radiated through an antenna, but mostly through cables and/or the mechanical structure of the system; therefore, its spatial radiation pattern is usually unknown but likely to be closer to isotropic than that of a directional antenna system; and iii) UEMR may have spectral contents which can be very variable depending on the type of electrical signals and design of the system. 

Telescopes used for radio astronomy normally receive UEMR from terrestrial sources located nearby (distances of kilometres) and predominantly though their sidelobes. There are many examples of radio telescopes dealing with terrestrial UEMR, as is the case of wind farms affecting LOFAR observations\footnote{\url{https://www.astron.nl/test-wind-turbine-near-lofar-meets-agreed-radio-emission-norms-2/}} or emission from microwaves resembling astrophysical signals \citep{petroff+15}. Radio astronomers also put great effort into shielding the necessary observation equipment (computers, receivers etc.) to avoid self-made UEMR to enter the data \citep{Swart+2}. Environmental interference (intended and unintended) to radio telescopes can be minimised by building them in designated radio quiet zones or RQZs (\citetalias{itu_ra2259_1}). Unfortunately, RQZs provide no mitigation against radio emission from Earth orbiting satellites, which radio telescopes can receive through their primary beam or near sidelobes. In the case of the Iridium satellite constellation, unwanted radio emissions (i.e. not UEMR) interfered with astronomical observations of the 1612~MHz OH spectral line for more than 20 years (e.g. \citealt{cohen04}, \citetalias{ecc_report_171}, \citetalias{ecc_report_247}). Studying reflections of terrestrial signals from satellites, \citet{prabu20} reported possible UEMR of two cubesats using the MWA between 80 and 103~MHz.

The proliferation of the new and large satellite constellations in low Earth orbit (LEO) -- often referred to as mega-constellations -- has caused worries in the astronomical community owing to the satellites ability to reflect sunlight and to emit radio signals \citep[e.g. ][]{2020ApJ...892L..36M,2020A&A...636A.121H,2021NatSR..1110642B}. This led to the Satellite Constellations workshops (SATCON1 and SATCON2; \citealt{walker+20_satcon1}), the Dark \& Quiet Skies I and II workshops \citep{walker+20_dqs1,walker+21_dqs2} and the founding of the IAU Centre for the Protection of the Dark and Quiet Skies from Satellite Constellation Interference (IAU CPS), the members of which investigated and continue to investigate the possible impact of large LEO satellite constellations on astronomy \citep[see][]{2020RNAAS...4..189R,bhg22,2023ivs..conf...19D}. Owing to the increasing total number of satellites in LEO, and hence the increasing probability that a satellite appears within the field of view of a radio telescope, it makes sense to consider satellite UEMR as a potential source of interference in the future. The potential threat posed by satellite UEMR from large constellations was first considered at the Dark \& Quiet Skies II workshop  \citep{walker+21_dqs2}.

In this paper we investigate the potential impact of satellite UEMR on radio astronomy through observations of the SpaceX Starlink satellite constellation. At the time of the observations presented here, this constellation was the largest in orbit with some 2100 satellites in orbit. This constellation provides broad-band internet connectivity with radio emission used for downlinks allocated to the 10.7 to 12.7~GHz frequency band\footnote{\url{https://fcc.report/IBFS/SAT-MOD-20200417-00037/2274316}}. Compatibility with radio astronomy observations in the protected 10.6$-$10.7~GHz band has previously been studied by the Electronic Communications Committee (ECC) of the European Conference of Postal and Telecommunications Agencies (CEPT) in its \citetalias{ecc_report_271}. As UEMR is predominantly expected at low frequencies (below $\sim1$~GHz) \citep[see][]{Pulkkinen}, well below the allocated radio transmission downlinks, we observed satellites belonging to the Starlink constellation at frequencies between 110 and 188~MHz with the LOFAR radio telescope \citep{hwg+13}.

This paper is organised as follows; Sect.~\ref{sec:UEMR_from_satellites} presents an overview of standards and regulations applicable to satellites and their subsystems, while Sect.~\ref{sec:impact_of_emr} uses simulations to investigate the potential aggregate impact of several satellite constellations and its maximum radiated power to comply with the ITU-R threshold levels in one of the protected radio astronomy bands. We describe the observations and their processing in Sect.~\ref{sec:observations_and_processing} and discuss the analysis of the detected signals in Sect.~\ref{sec:analysis}. Finally, Sect.~\ref{sec:summary} contains a summary and conclusions.

\section{UEMR of satellite systems}\label{sec:UEMR_from_satellites}
Typical satellites are composed of many different modules called subsystems, each one fulfilling a specific function for the satellite to operate. Satellite manufacturers make use of electromagnetic compatibility (EMC) to ensure that all the different subsystems will be compatible with each other. A typical EMC programme focuses on testing each subsystem to ensure that sufficient margins exist between emissions and susceptibilities for the ensemble to work without self-interference.

\begin{figure}[!tp]
    \centering\includegraphics[width=\columnwidth]{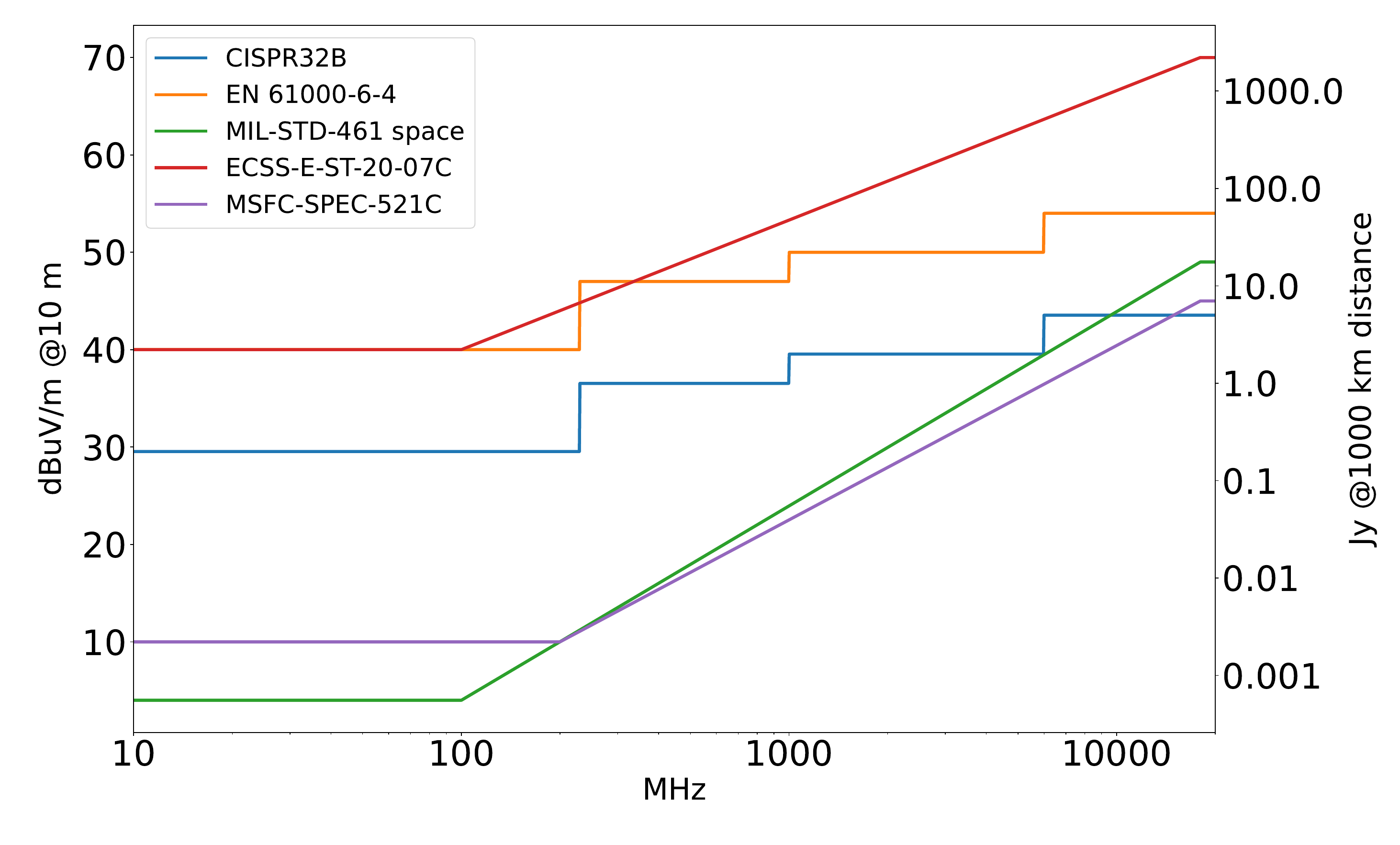}
    \caption{Radiated emission limits for several EMC standards such as commercial (CISPR, EN61000), military (MIL-STD-461), and space (MSFC and ECSS). Left axis shows electric field measured at 10~m distance, right axis shows equivalent spectral power flux density (in Jy) assuming a source at 1000~km distance.}
    \label{fig:Radiated_emission_standards}
\end{figure}

There are some EMC standards dedicated to space missions, such as the NASA MFSC-SPEC-521 or the ESA ECSS-E-ST-20-7C, most of them based on the US military standard MIL-STD-461. These EMC standards define, among other things, the maximum level of electromagnetic radiation that equipment can generate. Most standards for space are more stringent than the ones used for commercial apparatus such as CISPR-32 (see Fig. \ref{fig:Radiated_emission_standards}) but that is not a hard requirement, as a satellite does not need to be compatible with ordinary commercial equipment.

Once completely assembled, a satellite is usually characterised by a `system level' test that evaluates the overall UEMR (among many other parameters) of it as a whole. These tests can last for weeks, depending on the complexity of the satellite, making it a very expensive activity. For this reason, system level tests tend to focus on the minimum and necessary checks for each parameter of a complete satellite. A clear example of this can be seen in \citet{Blondeaux+3}, where UEMR is not highlighted as an important step to characterise a satellite constellation.

While commercial standards such as the IEC~61000 family, CISPR or the US Federal Communications Commission (FCC) part 15 (see Fig.~\ref{fig:Radiated_emission_standards}), are harmonised and mandatory to allow entry into a certain market, there is currently no international agency or space law that requires a spacecraft to comply to a certain EMC standard. Furthermore, the information about which EMC standard is used for a specific programme, the considered UEMR thresholds, or the real level of emissions are rarely made public. Few examples are in the public domain such as \citet{Yavas+1} and \citet{Elkman+2}. Informal communications with satellite industry specialists indicated that the normal practice for satellite level UEMR tests is to set an emission threshold relatively high (which speeds-up testing times) and only apply stringent levels (long testing times) to narrow frequency bands where the satellite or the rocket-launcher have receivers or sensitive instruments. In \citet{Yavas+1}, results of a satellite emission level test are shown, where the limit threshold (marked as a solid red line in their Fig.~7) is defined at very high levels of emission almost for every frequency with the exception of a few communication bands. 

Owing to this lack of information, we can suppose that a satellite could emit relatively strong UEMR signals, outside of the bands of interest for the manufacturer or operator, and still pass this type of testing. This is not an unlikely situation, since many subsystems can aggregate their emissions or their interconnection can change the electromagnetic configuration of the satellite and increase the emissions in a certain frequency band. This may not have been an issue in the past, with very small constellations or with single satellite systems. Even if a satellite had strong UEMR, it would require a very sensitive receiver to detect it or in other words would require the satellite to be in the main lobe of a radio telescope for a considerable fraction of an observation: a very rare condition until recently.

With the advent of the large LEO satellite constellations (such as Starlink phase 1 with 4408 satellites or OneWeb phase 1 with 720 satellites\footnote{\url{https://planet4589.org/space/con/conlist.html}.}) the situation changes. Firstly, the number of LEO satellites leads to an increase of the aggregate signal, which might become large enough to cause interference even through the sidelobes and increases the probability of a detection in the main lobes of the radio telescope. Secondly, the new satellites are manufactured in series, therefore it is possible that many satellites present similar UEMR. These two effects could make the situation for radio astronomy complicated, even in radio bands reserved to radio astronomy.

\section{Potential impact of satellite EMR on RAS}\label{sec:impact_of_emr}

To investigate the potential impact of satellite EMR on radio astronomical observations, it is possible to make use of the established methods that were developed by ITU-R for regular compatibility calculations of wanted and unwanted emissions. The ITU-R recommends to use the equivalent power flux density (EPFD) method (see \citetalias{itu_s1586_1,itu_m1583_1}). A satellite constellation is simulated over a given time range. The power received from each satellite can be calculated from the transmitted power, taking into account transmitter and receiver antenna gains and path propagation losses (e.g. line of sight losses, atmospheric attenuation) before it enters the radio astronomy receiver. The total aggregated power, which is the sum of all power contributions, can be then determined. Under the assumption of standardised characteristics of the receiving antenna, the received power can also be converted to the associated power flux density (PFD, known as total or integrated flux density in the radio astronomy community), which allows to conveniently compare it to PFD threshold levels that are defined in regulations for the protection of a victim station. An advantage of this conversion is that it makes a better comparison possible between different receiving stations, which usually have different antenna patterns and gains. For example, the RAS protection criteria (\citetalias{itu_ra769_2}; in Tables 1 and 2) are provided for an isotropic receiver, although in reality radio telescopes usually have very high forward gain.

\subsection{Assessing the aggregate impact of a satellite constellation}
In the following, the EPFD method is used to determine the potential impact of UEMR from different satellite constellations on radio astronomy observations. The EPFD method is widely used in spectrum management and is well documented in ITU-R documents. For convenience, a more detailed summary is provided in Annex~\ref{appendix:epfd}. Here, the basic steps are explained in a simplified form. To calculate the received power for one particular pointing direction of the receiver antenna and a certain satellite orbit configuration, the procedure is as follows.

In the first step the satellite positions (and transmitter antenna orientations) with respect to the observer are be determined for a number of time steps and for a given period of time. The required time resolution mostly depends on the satellite altitudes. For low-earth orbit (LEO) satellites the time resolution should be 1~s or less as the angular velocities are high. Then the link budget (path propagation losses as well as transmitter and receiver antenna gains) between satellites and observer are computed. As the satellites are not necessarily in the main beam of the radio telescope, the angular separation between the antenna pointing direction and the geometrical position of the satellites needs to be accounted for, which changes the effective receiver gain. Likewise, the observer will usually not be situated in the forward direction of the satellite antenna. Modern satellites are often equipped with active antennas that allow electronic beam-forming in real-time, such that the effectively transmitted power towards the observer can fluctuate strongly. It should be noted, however, that in the case of UEMR, given its nature, a high directivity is not expected to be reached and an isotropic transmitting antenna pattern is used hereafter as an approximation. After the link budget is calculated, all the individually received powers (from each satellite) are added, which yields the total aggregated power. Finally the total aggregated power received at the radio telescope is compared to the permitted threshold levels, for example defined in \citetalias{itu_ra769_2}. In this recommendation, the RAS protection levels are specified for an integration time of 2000~s, thus it is necessary to simulate the orbits over this time span.

The calculation is performed for a grid of sky cells (or telescope pointing directions) having approximately equal solid angles. This allows to analyse the spatial distribution of the contributed power levels. To assess statistical scatter, the whole simulation is repeated hundreds or thousands of times for different starting times and antenna pointings within the grid cells.
 
Often, the power flux density at the observer location (caused by the satellites) is transformed into the so-called equivalent power flux density (EPFD). This is the power flux density, which would need to be present in the boresight of a radio telescope to create the same power as the aggregated power from all satellites.  Annex~\ref{appendix:epfd} contains more details on this.

\subsection{EPFD and large satellite constellations}\label{subsec:epfd_simulation_setup}

For some of the large satellite constellations under construction, in particular SpaceX/Starlink and OneWeb, EPFD calculations were performed by the Electronic Communications Committee (ECC) of the European Conference of Postal and Telecommunications Administrations (CEPT) in its \citetalias{ecc_report_271}. In that report, the out-of-band emissions of the satellite downlinks in the RAS band at 10.60$-$10.70 GHz were analysed by means of this method.

To our knowledge, UEMR from large satellite constellations in operation has never been studied nor measured, probably because the number of satellites (of the same design) was not large enough to even be considered a problem, but this situation has changed now. Using the EPFD method it is possible to determine the maximum UEMR that each single satellite of a constellation may radiate in the 150.05$-$153~MHz primary radio astronomy band, while not producing harmful interference. Here we consider harmful interference as defined in \citetalias{itu_ra769_2}. 

The 150.05$-$153~MHz frequency band, which is allocated to the RAS, was chosen as it is commonly accepted that radiation caused by electronic circuits is mainly concentrated below 1~GHz, and it falls within the observing band of LOFAR. The harmful interference threshold in this band is $-194~\mathrm{dB}\left[\mathrm{W}\,\mathrm{m}^{-2}\right]$ over a bandwidth of about 3~MHz, according to \citetalias{itu_ra769_2} (see their Table 1).

Given that the actually radiated emissions from a single satellite are unknown, we have to assume some value. An electric field strength of $30~\mathrm{dB}\left[\mathrm{\mu V}\,\mathrm{m}^{-1}\right]$ is a typical radiation level\footnote{The value of $30~\mathrm{dB}\left[\mathrm{\mu V}\,\mathrm{m}^{-1}\right]$ in CISPR refers to a so-called quasi-peak detector. Here, we assume that field strength to be the average over the measurement period. Usually, the two detector types may lead to significantly different outputs -- by several dB -- depending on the properties of the signal \citep[compare][and references therein]{winkel19}} found in commercial standards such as CISPR-32 based on a detector bandwidth of 120~kHz and measured at a distance of 10~m. This number is equivalent to a radiated spectral power of $-45.6~\mathrm{dB}\left[\mathrm{mW}\,\mathrm{MHz}^{-1}\right]$. We also assume in our simulations that this radiation is constant in time and frequency within the studied band. In practice this is certainly not the case. UEMR features can be time-variable and could also be narrow-band and in such a case a bandwidth correction factor would need to be applied. We furthermore work under the simplification that satellite UEMR is isotropically radiated.

The RAS antenna pattern and gain used in the calculations depends on the type of radio telescope. At these low frequencies, mostly interferometric telescopes are used, such as LOFAR and SKA1-Low. The actual antenna patterns of interferometers (after beam-forming and correlation) are complex and are not perfectly described by the \citetalias{itu_ra1631_0} model. Therefore, we perform the EPFD assuming parabolic-dish antennas of diameter 25-m and 70-m, respectively, which approximately have the same effective antenna area as SKA1-Low tiles and LOFAR (international) stations. In our simulations it is assumed that the RAS station is located at the geographical latitude of LOFAR, $53^\circ$\,N.

Using these parameters and assumptions, EPFD calculations were carried out for a number of existing or currently in-deployment satellite constellations: Spire\footnote{\url{https://fcc.report/IBFS/SAT-LOA-20151123-00078/1126653}}, Iridium NEXT\footnote{\url{https://fcc.report/IBFS/SAT-AMD-20151022-00074/1145619}}, OneWeb\footnote{\url{https://fcc.report/IBFS/SAT-LOI-20160428-00041/1135071}}, SpaceX/Starlink\footnote{\url{https://fcc.report/IBFS/SAT-MOD-20200417-00037/2274316}}, and SpaceX/Swarm\footnote{\url{https://fcc.report/IBFS/SAT-LOA-20181221-00094/1592875}}. This provides us a range of constellation sizes from 66 satellites up to 4408 (see Tab.~\ref{tab:epfd_results}) in various orbital configurations. For the satellite position calculations we made use of the open-source Python package \texttt{cysgp4}\footnote{\url{https://pypi.org/project/cysgp4/}} \citep{cysgp4}, which is available under GPL-v3 license. It is a wrapper around the \texttt{sgp4lib}\footnote{\url{https://github.com/dnwrnr/sgp4}} C++ implementation of the simplified perturbation model SGP4 \citep[see also][]{vallado06}. Furthermore, the \texttt{pycraf}\footnote{\url{https://pypi.org/project/pycraf/}} Python package \citep{pycraf,winkel18} was used, which provides implementations for a number of relevant ITU-R Recommendations. It is also available under GPL-v3 license.

\subsection{Simulation results}\label{subsec:epfd_simulation_results}

\begin{figure}[!tp]
    \centering\includegraphics[width=\columnwidth]{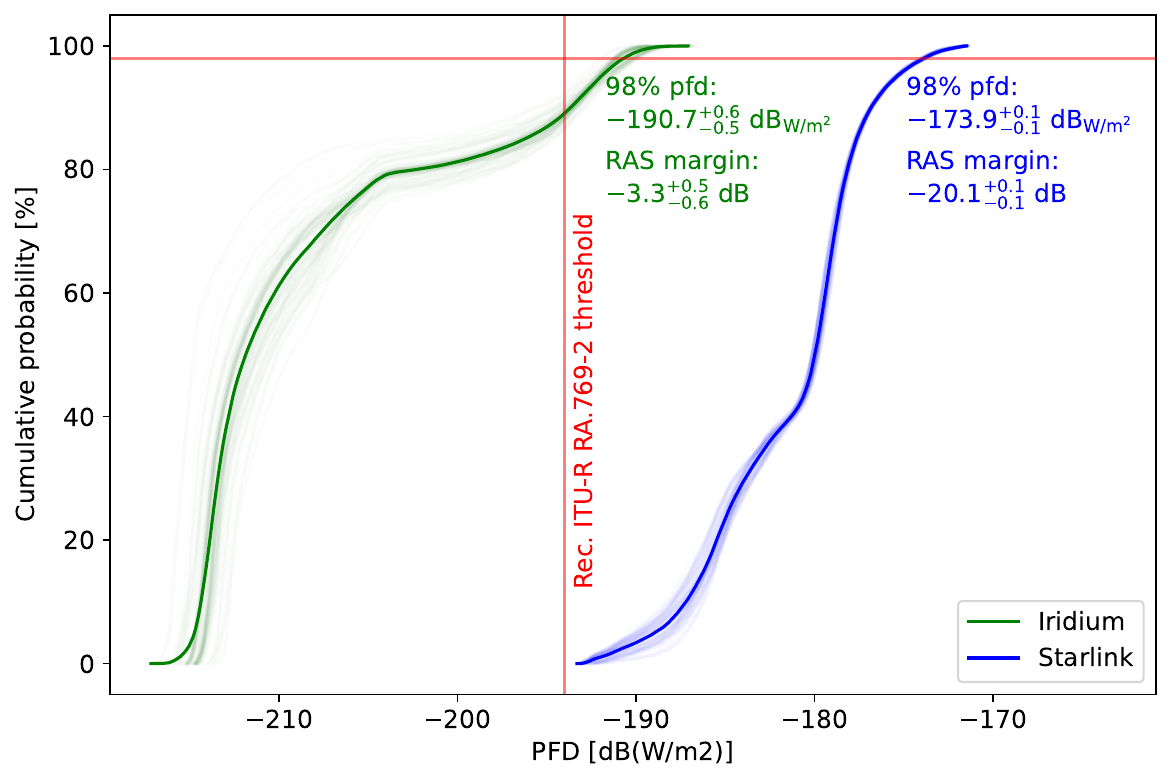}
    \caption{Cumulative distribution functions for EPFD values owing to Iridium NEXT and Starlink, assuming an isotropic transmitter spectral power of $-45.6~\mathrm{dB}\left(\mathrm{mW}\,\mathrm{MHz}^{-1}\right)$ and a 70-m radio telescope located a geographic latitude of $53^\circ$\,N.}
    \label{fig:emi_cumulative_data_loss_horizontal}
\end{figure}

\begin{figure}[!tp]
    \centering\includegraphics[width=\columnwidth]{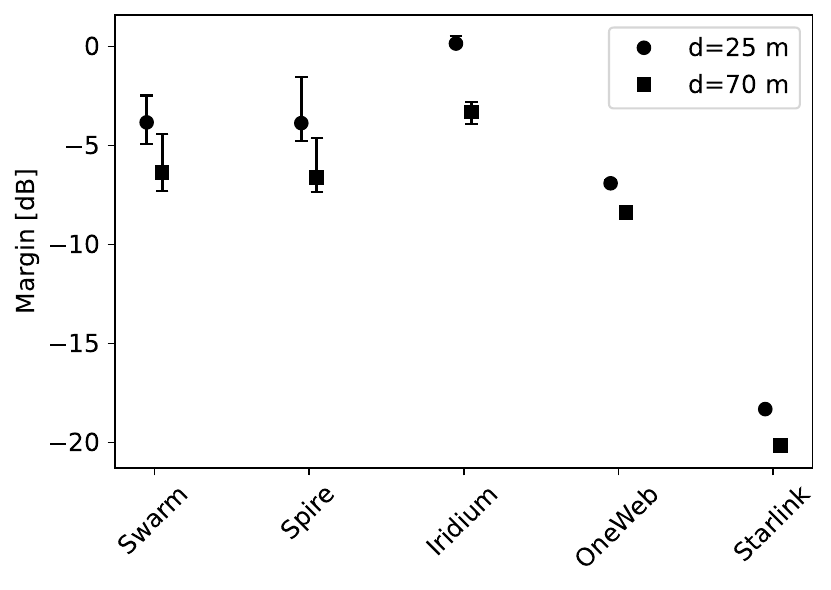}
    \caption{Calculated margins for all simulated satellite constellations with the assumption of a $30~\mathrm{dB}\left[\mathrm{\mu V}\,\mathrm{m}^{-1}\right]$ UEMR with respect to the ITU-R thresholds in 
    150.05$-$153~MHz.}
    \label{fig:epfd_all_margins}
\end{figure}

For each constellation in Table~\ref{tab:epfd_results}, one hundred iterations (simulation runs) were processed, which allows us to assess the statistical scatter of the results. As an example for the results, Fig.~\ref{fig:emi_cumulative_data_loss_horizontal} shows the cumulative distribution function for EPFD values for the Iridium NEXT and Starlink constellations with the assumption of UMR with an electric field strength of $30~\mathrm{dB}\left[\mathrm{\mu V}\,\mathrm{m}^{-1}\right]$ over the full RAS bandwidth\footnote{If there was only a narrow-band signal within the RAS band, a correction factor would need to be applied.} and a RAS antenna with a 70-m diameter located a geographic latitude of $53^\circ$\,N. The light green and blue curves in the figure show the results for all sky cells in each individual simulation run, while the darker curves represent the median of the individual runs in each sky cell. \citetalias{itu_ra1513_2} recommends that the total data loss caused by a single interfering system should not exceed 2\%, which is indicated by the horizontal red line (the 98\% percentile) in the figure. The vertical red line marks the \citetalias{itu_ra769_2} threshold. The cumulative probability at which this threshold is exceeded can be used to determine the actual expected data loss (about 10\% for Iridium NEXT and 100\% for Starlink with the assumptions used in the simulation). The intersection between the cumulative probability curve and the horizontal red line of 98\% percentile yields the so-called margin, that is the difference between the RAS threshold and the actual received power flux density. If it is negative, emissions from the respective satellite constellation ought be below  the assumed model values by that amount in order to comply with the thresholds in the RAS band. The inferred margins for all satellite constellations are presented in Fig.~\ref{fig:epfd_all_margins}.

Based on the margins, under the assumptions used in the simulation, it is possible to determine a maximum electric field value that each satellite should comply with to ensure that the received power at the RAS station is not in excess of the permitted RAS threshold levels at the data loss of 2\%. These values are summarised in Tab.~\ref{tab:epfd_results}. It is noted that the calculated values are lower than commercial EMC standard thresholds such as the CISPR-32 Class B with $30~\mathrm{dB}\left[\mathrm{\mu V}\,\mathrm{m}^{-1}\right]$.

\begin{table}[!tp]
\caption{Results of the EPFD simulations: to avoid exceeding the RAS threshold levels in the band 150.05$-$153~MHz, the electric field values produced by satellites (measured with a 120~kHz detector at 10~m distance) should be lower than the stated E-fields. This assumes that the UEMR has broad-band type, or more precisely that the signal is constant over the full RAS band, 150.05$-$153~MHz. With a measurement bandwidth of 2.95~MHz the electric field limits are $(2950/120)=13.9~\mathrm{dB}$ higher.}
\label{tab:epfd_results}
\centering
\begin{tabular}{l r r l l}
\hline\hline
\rule{0ex}{3ex}Satellite System & \# Sats & Avg. alt. & \multicolumn{2}{c}{Max. E-field}\\
  &  & km & \multicolumn{2}{c}{$\mathrm{dB}\left[\mathrm{\mu V}\,\mathrm{m}^{-1}\right]$}\\[1ex]
\hline
\rule{0ex}{3ex} &  &  & $d=25~\mathrm{m}$& $d=70~\mathrm{m}$\\
\hline
\rule{0ex}{3ex}Spire &  118 &  500 & $26.1_{-0.9}^{+2.3}$ & $23.4_{-0.7}^{+2.0}$\\
\rule{0ex}{3ex}Iridium NEXT         &   66 &  780 & $30.1_{-0.2}^{+0.4}$ & $26.7_{-0.6}^{+0.5}$\\
\rule{0ex}{3ex}OneWeb               &  720 & 1200 & $23.1_{-0.1}^{+0.2}$ & $21.6_{-0.2}^{+0.2}$\\
\rule{0ex}{3ex}SpaceX/Starlink      & 4408 &  560 & $11.7_{-0.1}^{+0.1}$ &  $9.9_{-0.1}^{+0.1}$\\
\rule{0ex}{3ex}SpaceX/Swarm         &  150 &  500 & $26.2_{-1.1}^{+1.4}$ & $23.6_{-0.9}^{+1.9}$\\[1ex]\hline
\end{tabular}
\end{table}

\begin{figure}[!tp]
    \centering
    \includegraphics[width=\columnwidth]{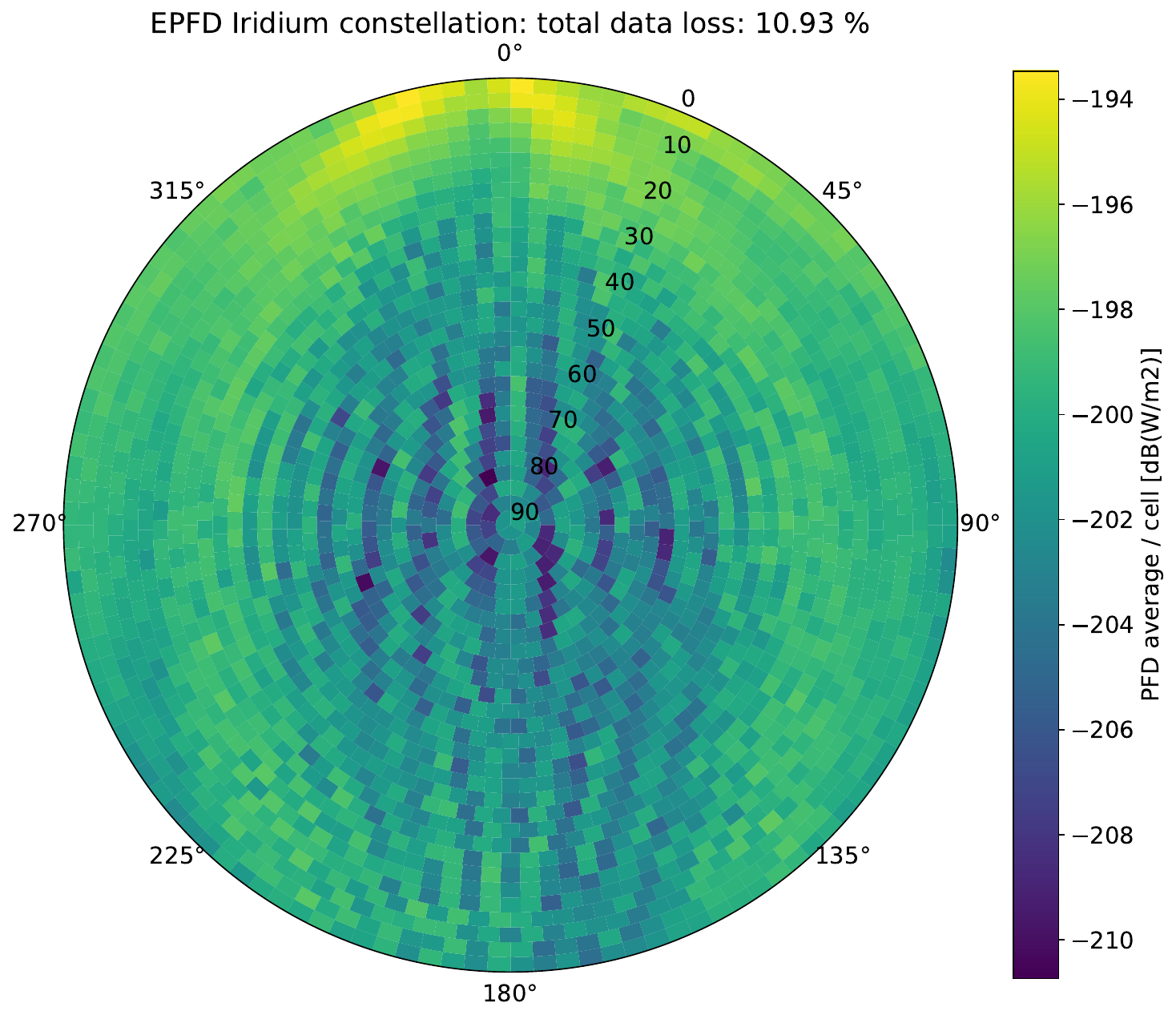}
    \includegraphics[width=\columnwidth]{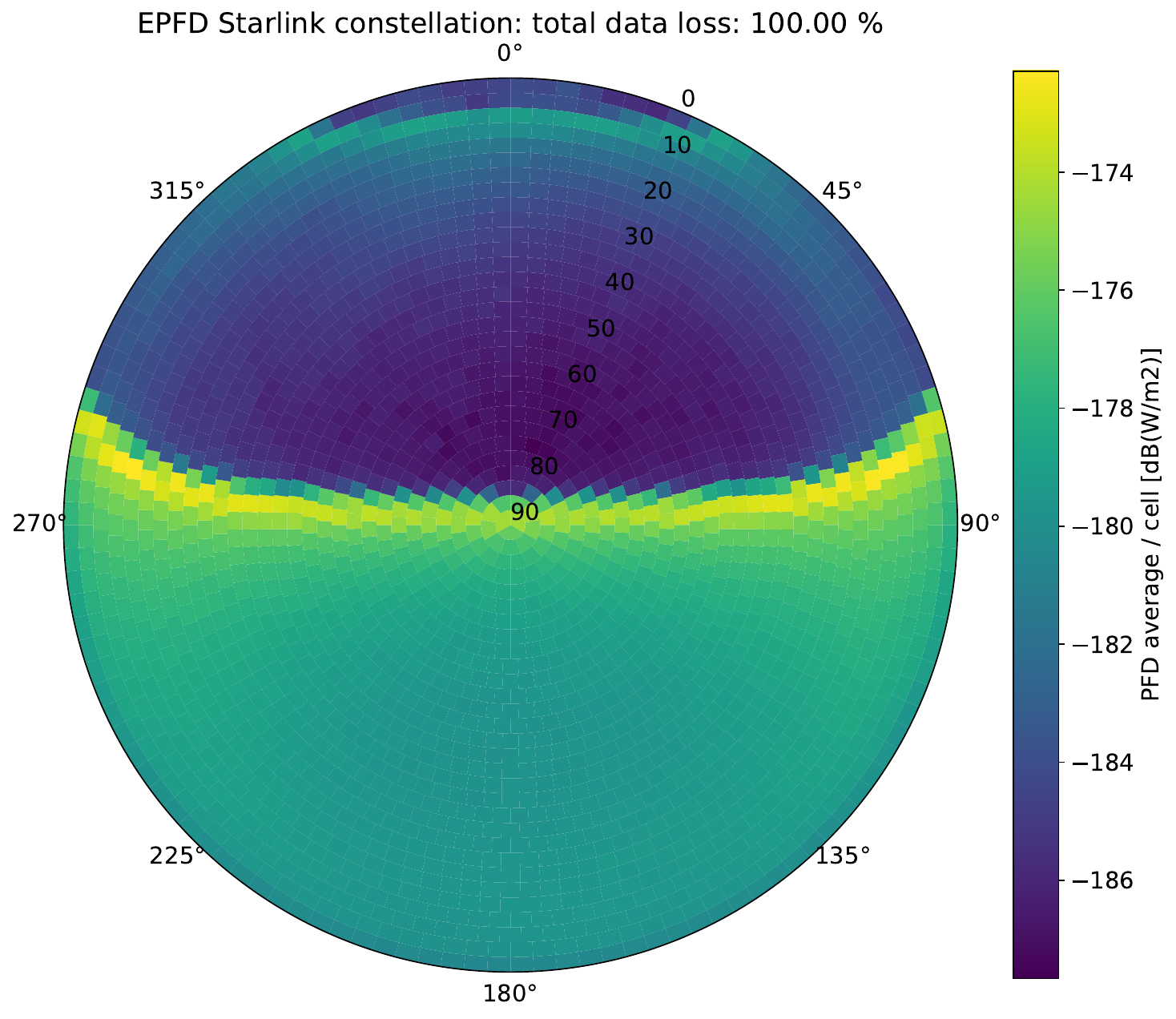}
    \caption{EPFD received in each sky cell (average over 100 iterations) owing to Iridium NEXT and Starlink constellations in topocentric frame (azimuth and elevation) as received by a 70-m RAS antenna.}
    \label{fig:epfd_starlink_iridium_skycells}
\end{figure}

It is also possible to investigate the regions on the visible (topocentric) sky, which contribute most to the overall received flux density, see Fig.~\ref{fig:epfd_starlink_iridium_skycells}, which shows the average EPFD per sky grid cell for the Iridium NEXT and Starlink constellations assuming a 70-m RAS antenna.

\section{Observations, data calibration and signal detection}\label{sec:observations_and_processing}
Based on the results obtained in Sec.~\ref{sec:impact_of_emr}, especially the ones for large satellite constellations such as Starlink, we conducted an observation with the LOFAR telescope which not only covers the frequency range of interest but can also produce multiple beams simultaneously increasing the probability of detecting satellite emissions within a reasonably short campaign. This section describes the observation method, data calibration and processing, and different types of detected signals.

\subsection{Observations}
\begin{figure}[!tp]
    \includegraphics[width=\columnwidth]{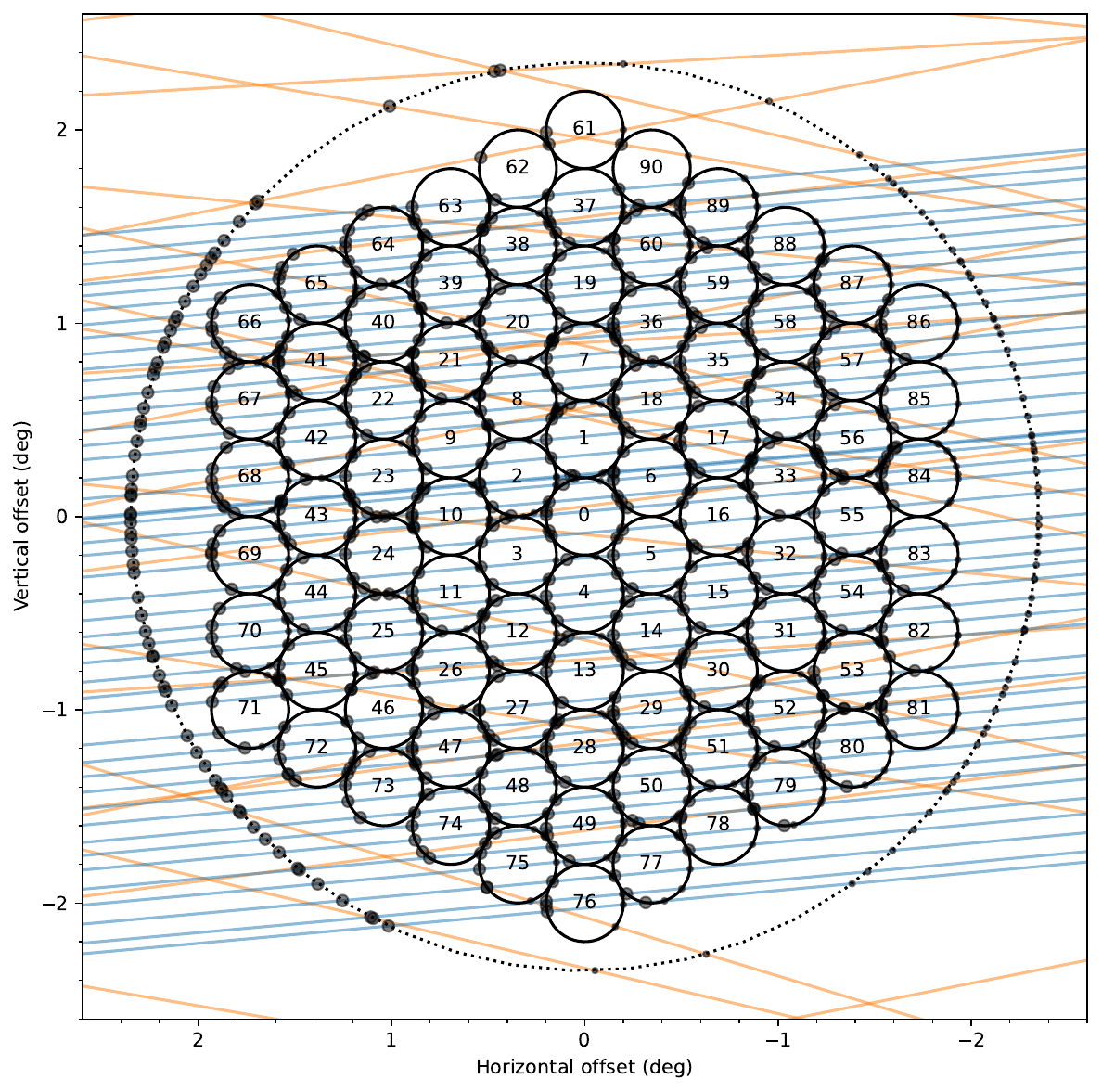}
    \caption{Beam pattern of the LOFAR observation in equatorial coordinates (right ascension and declination). The 91 tied-array beams are indicated with the smaller circles ($24\arcmin$ FWHM at 150~MHz), while the larger dashed circle denotes the FWHM of a LOFAR core station ($4\fdg7$ at 150~MHz). Predictions of the motion of Starlink satellites with respect to the beam pattern are indicated with the blue (at orbital altitude $h=358$~km) and orange (at $h=550$~km). The small and large black circles indicate the ingress into and egress from the tied-array and station beams, respectively.}
    \label{fig:beam_layout}
\end{figure}

LOFAR, the Low Frequency Array \citep{hwg+13}, is a network of telescopes with stations spread over Europe and a dense core in the north of the Netherlands. We obtained a 1-hour observation targeting mostly SpaceX/Starlink satellites on 2022 April 1, starting at 18:30:00\,UTC. Radio signals from the High Band Antennas (HBA) of the central six LOFAR core stations, those on the Superterp, were coherently beam-formed by the \textsc{Cobalt} beam-former \citep{bmn+18} to form 91 tied-array beams (TABs). The TABs were distributed in five hexagonal rings covering the $4\fdg7$ full width at half maximum (FWHM) station beam, each ring separated by $24\arcmin$ from the next; see Fig.~\ref{fig:beam_layout}. This separation was chosen such that the TABs overlap at the half-power point around 150~MHz, assuming circular beams with a $24\arcmin$ FWHM at 150~MHz. For each tied-array beam, (uncalibrated) Stokes~I intensities in the form of dynamic spectra were recorded between 110 and 188~MHz, with 10.48~ms time resolution and 12.21~kHz frequency resolution.

The TABs were centred towards, and tracking, $\alpha_\mathrm{J2000}=08^\mathrm{h}00^\mathrm{m}00^\mathrm{s}$ and $\delta_\mathrm{J2000}=+49\degr30\arcmin00\arcsec$. This pointing direction was chosen for its high Galactic latitude ($b=31\fdg1$ at Galactic longitude $l=169\fdg4$) and hence low sky temperature (reducing the overall system temperature), as well as the high elevation above the horizon of LOFAR (maximum elevation of $86\fdg5$ at 18:54UTC), minimising the range between Starlink satellites at their operational altitude of 550~km. Furthermore, at the latitude of LOFAR ($\phi=52\fdg92$), the currently most populated Starlink shells (with orbital inclinations of $53\fdg0$ and $53\fdg2$) lead to over-densities of satellites per unit area of sky near LOFAR's zenith \citep{bhg22,lbr22}, maximising the number of Starlink satellites passing through the TABs.

We used public ephemerides\footnote{Distributed through \url{www.space-track.org}.} of the Starlink satellites generated by SpaceX for the observation planning and the processing of the data. The public ephemerides provide predictions for position and velocity of each Starlink satellite with respect to an Earth-centred inertial coordinate frame at 1~min time intervals, and include planned manoeuvres to adjust the satellite orbit. From these ephemerides, the trajectory of each satellite passing through the LOFAR beam pattern during the 1~hour observation was calculated, resulting in the passes shown in Fig.~\ref{fig:beam_layout}. We also computed the time of ingress and egress of each satellite through the station beam and the TABs. We note that individual Starlink satellites are known to make small unplanned manoeuvres, which generally result in the satellite passing early or late compared to predictions, without significantly altering its trajectory on the sky.  The ephemerides show that a total of 68 individual Starlink satellites passed through the LOFAR station beam during the 1~hour observation, 22 of which were at the operational altitude of $h=550$~km. The other 46 Starlink satellites passing through the beam pattern were at an altitude of 350~km. These satellites belonged to a group of 48 satellites launched on 2022 March 9, 23~days before our observations, and were still raising their orbits to operational altitudes. The Starlink satellites of this launch are of a newer version 1.5 type\footnote{\url{space.skyrocket.de/doc_sdat/starlink-v1-5.htm}} compared to the Starlink satellites at the operational altitudes, which reportedly are version 1.0. 

The properties of these satellites and their passes through the beam pattern are provided in Table~\ref{tab:satellite_passes}. Owing to the high elevation of the observations above the horizon, the distances to the Starlink satellites in the operational orbits at 550~km was around 555~km, while the orbit raising group were at distances of around 356~km. These distances are in the far field of the LOFAR Superterp, whose maximum baseline of $\sim$300~m puts the Fraunhofer distance from 66 to 113~km for the observed LOFAR band of 110 to 188~MHz. At these distances, the satellites crossed the $4\fdg7$ FWHM of the station beam within 6 and 4~s, respectively, while the $24\arcmin$ TABs were crossed within 0.54~s for satellites at 550~km altitude, and 0.34~s for those at 350~km altitude ($t_\mathrm{pass}$ column of Table~\ref{tab:satellite_passes}). Of the 68 satellite passes, only two did not pass through any of the TABs, while the majority of the others passed through several adjacent TABs, as indicated by the $n_\mathrm{TAB}$ column in Table~\ref{tab:satellite_passes}. Finally, we note that all Starlink satellites passing through the beam pattern during this observation were illuminated by the Sun, and that their solar panels could have been generating power.

\subsection{Data calibration}
To calibrate the recorded data-sets, we performed both the frequency-dependent system gain (band-pass) correction\footnote{It is noted that the \textsc{Cobalt} beam-former applies a band-pass correction for every single spectral sub-band to correct for the digital filter curve of the poly-phase filter-bank.} as well as the intensity calibration. 
For a single dish antenna, the on-source, off-source method represents a useful strategy to correct for the system gain. In very simple terms, the recorded uncalibrated power spectrum, $P(t_i, f_j)$, at time, $t_i$ and in frequency channel $f_j$ is related to the actual antenna temperature, $T_A$, via the receiver system transfer function, $G_\mathrm{bp}(t_i, f_j)$. $G_\mathrm{bp}$ is a function of frequency, but it also depends mildly on $t_i$ owing to slow drifts of the receiver (amplifier) gain. For the accuracy required for this project, one can safely assume that $G_\mathrm{bp}$ is constant with time over the relatively short observation period. Thus,
\begin{equation}
P(t_i, f_j) = G_\mathrm{bp}(f_j) T_A(t_i, f_j)\,.
\end{equation}

The idea of the on-source, off-source method is to divide two spectra to remove the frequency-dependent band-pass shape \citep[compare][]{winkel12}. This yields
\begin{equation}
\frac{T_\mathrm{source}}{T_\mathrm{sys}}=\frac{P^\mathrm{on}}{P^\mathrm{off}} - 1\,,\label{eq:simple-onoff}
\end{equation}
where it was assumed that $T_A^\mathrm{on} = T_\mathrm{source} + T_\mathrm{sys}$, while $T_A^\mathrm{off} = T_\mathrm{sys}$. The quantity $T_\mathrm{source}$ denotes the signal from a source to be measured, which would only be in the on-source spectrum, while all other constituents to the antenna temperature are denoted as system temperature, $T_\mathrm{sys}$. Of course, anthropogenic signals, which are often highly variable with time and frequency, would produce residual imprints in the resulting data and ideally needs to be treated before the method is applied. Furthermore, any astronomical signal that is present in both the on- and off-source observation (e.g. large-scale continuum radiation) would also not be processed properly by the method. 

Classically, the on-source, off-source strategy involves position switching as one needs a measurement without the (astronomical) source of interest for the reasons explained above. However, LEO satellites are within the observation beam for a very short amount of time, only. Thus, the off-source spectrum can simply be constructed by choosing data at a different time, for example shortly before and after a satellite crosses the beam, and taking the average spectrum over this time range. Another possibility would be to determine the off-source spectrum over the full time span of the observation, for example by averaging all spectra leaving out those that are associated with satellite crossings. The second method should only be applied, though, if the temporal stability of $G_\mathrm{bp}$ is sufficient. Here, both strategies have been tried out and no significant difference in the calibrated data-sets was found. In practice, all the averaging steps in the above procedures could also make use of the median estimator, which is more robust against outliers, produced by short-term anthropogenic signals. 

Obviously, the beam-formed LOFAR data is not measured with only a single antenna. Nevertheless, the method outlined above can still be used in a very similar manner. The measured power spectrum, $P$, is again subject to a frequency dependent `system gain', which is now acting on an `effective (ensemble) antenna temperature' instead of each element's antenna temperature. The on-source, off-source method will remove the imprint of this system gain from the data, but the resulting quantity is not simply $T_\mathrm{source}/T_\mathrm{sys}$ as in Eq.~\ref{eq:simple-onoff} but a different quantity.

For the absolute flux calibration we used the approach outlined in \citet{kondratiev16} which models the effective area, beam shape, system temperature and coherence of LOFAR. The radiometer equation \citep[e.g.][]{dewey+85} relates the (power) flux density root mean square (RMS) at TAB level to these quantities by
\begin{equation}
    \Delta S_\nu^\mathrm{tab} = \frac{\Delta T^\mathrm{station}}{\Gamma_\mathrm{tab}} =  \frac{1}{\Gamma_\mathrm{tab}}\frac{T_\mathrm{sys}^\mathrm{station}}{\sqrt{n_\mathrm{p} t_\mathrm{obs} \Delta\nu}}\,,
\end{equation}
where $\Delta T^\mathrm{station}$ is the noise level that can be achieved with a single station based on the radiometer equation. It depends on the system temperature of a station, $T_\mathrm{sys}^\mathrm{station}$, the number of polarisation channels, $n_\mathrm{p}=2$, that were averaged, the integration time, $t_\mathrm{obs}$, and the bandwidth, $\Delta\nu$, which in this case is the width of a spectral channel. The quantity $\Gamma_\mathrm{tab}=0.5A_\mathrm{eff}^\mathrm{tab}/k_\mathrm{B}$ is the sensitivity or gain that translates between the station level system noise and the TAB flux density RMS. It is determined by the effective aperture area of a TAB, $A_\mathrm{eff}^\mathrm{tab}$, and the Boltzmann constant $k_\mathrm{B}$. The value of $A_\mathrm{eff}^\mathrm{tab}$ depends on the beam-forming efficiency and the number of contributing antennas. \citet{kondratiev16} derived an approximation formula,
\begin{equation}
    A_\mathrm{eff}^\mathrm{tab} = \eta_\mathrm{active} N^{0.85}A_\mathrm{eff}^\mathrm{station}\,,
\end{equation}
with the fraction of active dipoles, $\eta_\mathrm{active}=0.95$ (about 5\% of the dipoles are typically not in operation), the number of HBA sub-stations in the Superterp, $N=12$, and the effective aperture area, $A_\mathrm{eff}^\mathrm{station}$, of one of these sub-stations. \citet{hwg+13} report on the values of $A_\mathrm{eff}^\mathrm{station}$ for a number of frequencies. For the frequencies used in this paper, we interpolated these values linearly. \citet{hwg+13} also estimated the system equivalent flux density (SEFD), which is the equivalent of the $T_\mathrm{sys}^\mathrm{station}$ on the flux density scale. The SEFD is relatively constant in the frequency range considered in the following, with a value of about 3~kJy. This can be converted to the system temperature scale using $\mathrm{SEFD}~[\mathrm{Jy}]=2760~ T_\mathrm{sys}^\mathrm{station}~\left[\mathrm{K}\right] / A_\mathrm{eff}^\mathrm{station}~\left[\mathrm{m}^2\right]$ \citep{hwg+13}.

Based on these equations and previously reported quantities, the calibration parameters in Table~\ref{tab:calibration_parameters} were determined for use in the subsequent sections. Because the station aperture, $A_\mathrm{eff}^\mathrm{station}$, appears in both terms, $T_\mathrm{sys}^\mathrm{station}$ and $\Gamma_\mathrm{tab}$, $\Delta S_\nu^\mathrm{tab}$ is actually independent on $A_\mathrm{eff}^\mathrm{station}$. It has a value of 2.986~Jy for all frequencies in Tab.~\ref{tab:calibration_parameters} (a flat SEFD was assumed). In order to calibrate the spectra, it is  only necessary to determine the noise level (in arbitrary units) and scale the data such that its RMS equals $\Delta S_\nu^\mathrm{tab}$.

\begin{table}[!tp]
\caption{Calibration parameters as explained in the text.}
\label{tab:calibration_parameters}
\centering
\begin{tabular}{l l l l l}
\hline\hline
\rule{0ex}{3ex}Frequency & $A_\mathrm{eff}$ & $A_\mathrm{eff}$ & $T_\mathrm{sys}$ & $\Gamma$ \\
          & Station        & TAB            & Station  & TAB\\
          
[MHz]     & [m$^2$]          & [m$^2$]          & [K] & [K/Jy] \\
\hline
\rule{0ex}{3ex}125  &  585 & 4597 & 636 & 1.66 \\
150  &  512 & 4021 & 557 & 1.46 \\
160  &  460 & 3611 & 500 & 1.31 \\
175  &  382 & 2997 & 415 & 1.09 \\
\hline
\end{tabular}
\end{table}

\subsection{Signal detection}
Any radio emission associated with Starlink satellites is expected to coincide in time with the predicted passage of a satellite through the LOFAR TABs, though it is a priori unclear if the radio emission would be broad-band, narrow-band, or a combination of both. The search is made difficult, however, as in the LOFAR observing band a lot of active radio services are operated producing signals which could by chance also appear at the same time when a satellite is predicted to pass through the beam.

The band from 110 to 188~MHz under consideration is allocated to several radio services such as  air traffic control (118--137, 138--144~MHz), amateur radio (144--146~MHz), emergency pagers (169--170~MHz), satellite transmissions (137--138, 148--150~MHz) and digital audio broadcasting (174--230~MHz), with emergency pagers and digital audio broadcasting being the strongest sources of radio emission\citep{offringa13}. The majority of these emission sources are terrestrial and hence are located close to, or on the horizon. As such, these signals will be detected in the sidelobes of the LOFAR station beam and TABs, and hence will appear at the same time and with similar signal strength in all TABs. On the contrary, objects moving through the sky will produce signals in the dynamic spectra of the TABs at different times as they pass through the TABs. This not only applies to the target Starlink satellites, but also to other satellites as well as aircraft.

Based on these considerations, the data-set has been independently searched for signals to avoid biases. We first found a narrow-band signal at 175~MHz and broad-band features with varying intensity spread across the band. Different data processing strategies were applied for this, which apparently were suitable to find the two types of signals. After the first detections were made, it also became clear that some of the satellite positions were not accurately predicted by the ephemerides. However, these first finding made it possible to correct the positional data, which triggered additional detections at further frequencies. In the following we provide a summary of the process.

\begin{figure*}[!tp]
    \includegraphics[width=\textwidth]{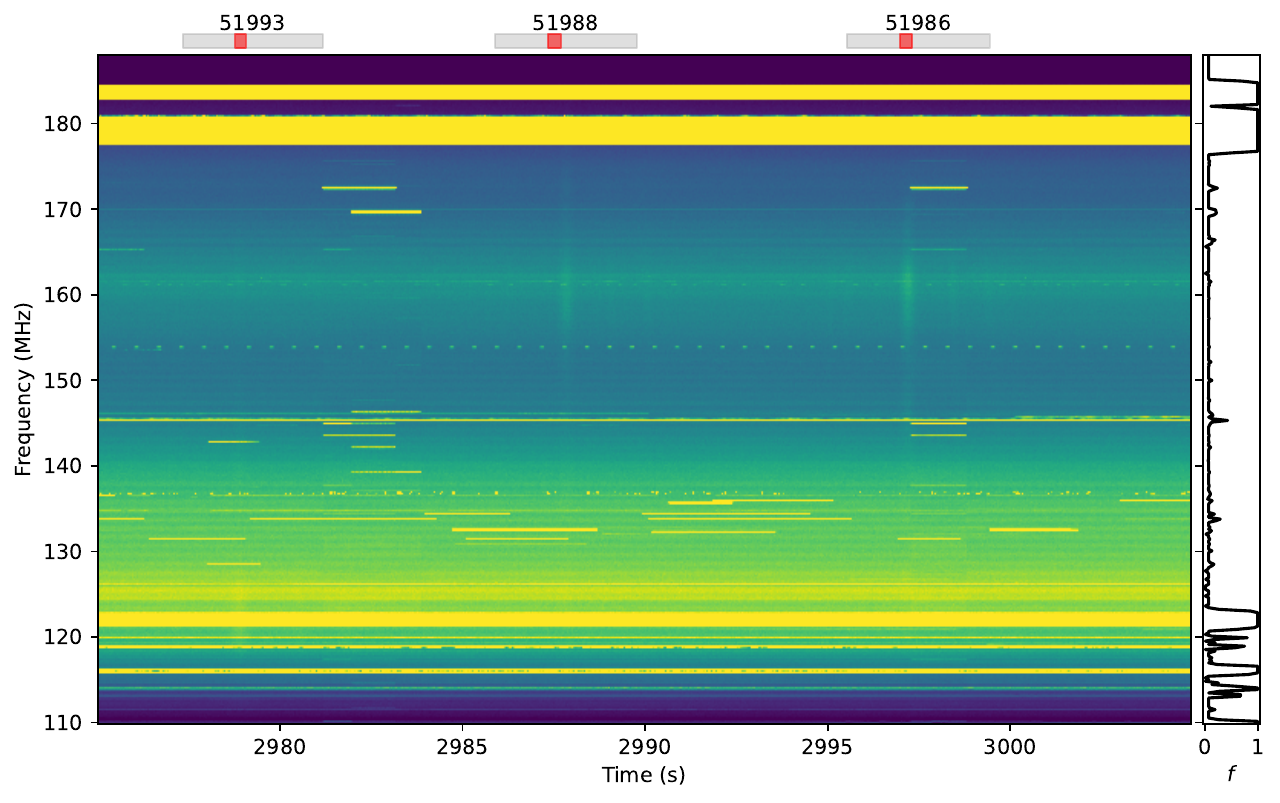}
    \caption{Dynamic spectrum of tied-array beam 18, showing broad-band radio emission of three Starlink satellites (NORAD IDs 51993, 51988 and 51986) coincident with the predictions from satellite ephemerides. For NORAD ID 51993, the emission is visible from 115 to 130~MHz, while objects 51988 and 51986 are more obvious from 140 to 175~MHz. The dynamic spectrum has been averaged by a factor four in time to a time resolution of 41~ms, and a factor 16 in frequency to a frequency resolution of 0.195~MHz. To show the temporal and spectral structure of the satellite emission, as well as that of other anthropogenic signals, the raw, uncalibrated dynamic spectrum is shown, without masking of anthropogenic signals. The bars at the top of the dynamic spectrum indicate the predicted time ranges where the indicated satellite passed through the LOFAR station beam (in grey), and the specific tied-array beam (in red). In the case of object 51988, the emission is about 0.33~s delayed compared to the prediction. The histogram on the right shows the fraction of the dynamic spectra that would have been masked in frequency by \texttt{OAflagger} \citep{offringa+12}.}
    \label{fig:raw_beam18}
\end{figure*}

Most of the brighter broad-band signals were already visible in the raw, uncalibrated dynamic spectra of the TABs after binning; see Figure~\ref{fig:raw_beam18}. As the duration of a pass through a TAB is of order 0.1 to 0.6~s, the dynamic spectra were averaged to a time resolution of 41.94~ms, keeping the frequency resolution fixed to 12.21~kHz. Next, \texttt{AOflagger} \citep{offringa+12} was used with the standard LOFAR flagging strategy to identify non-astrophysical signals and create a mask for the dynamic spectrum of each TAB. We found that, on average, 23\% of the dynamic spectrum is flagged, 6.25\% of which is due to each 16th channel, which contains the DC component of 16~channel poly-phase filter-bank used to channelise the LOFAR 0.195~MHz sub-bands into 12.21~kHz channels.

For each Starlink satellite passing through the LOFAR station beam, we started by extracting 20~s in time centred on the predicted mid-point of the pass through the LOFAR station beam from each of the 91 TABs. For this we used the band-pass calibrated dynamic spectra. To minimise the impact of terrestrial signals, which often appear similar in all beams, we subtracted from the extracted dynamic spectrum of each TAB the mean of the dynamic spectra of all the other TABs. Finally, again for each satellite pass, the resulting dynamic spectra of those TABs through which the satellite passed were aligned in time based on the predicted passage time and averaged to increase the signal-to-noise of any satellite emission. We note that with this approach we specifically chose not to mask any data that was flagged by \texttt{AOflagger}, this was to ensure that no emission from satellites would be removed from the analysis.

\begin{figure*}[!tp]
  \includegraphics[width=\textwidth]{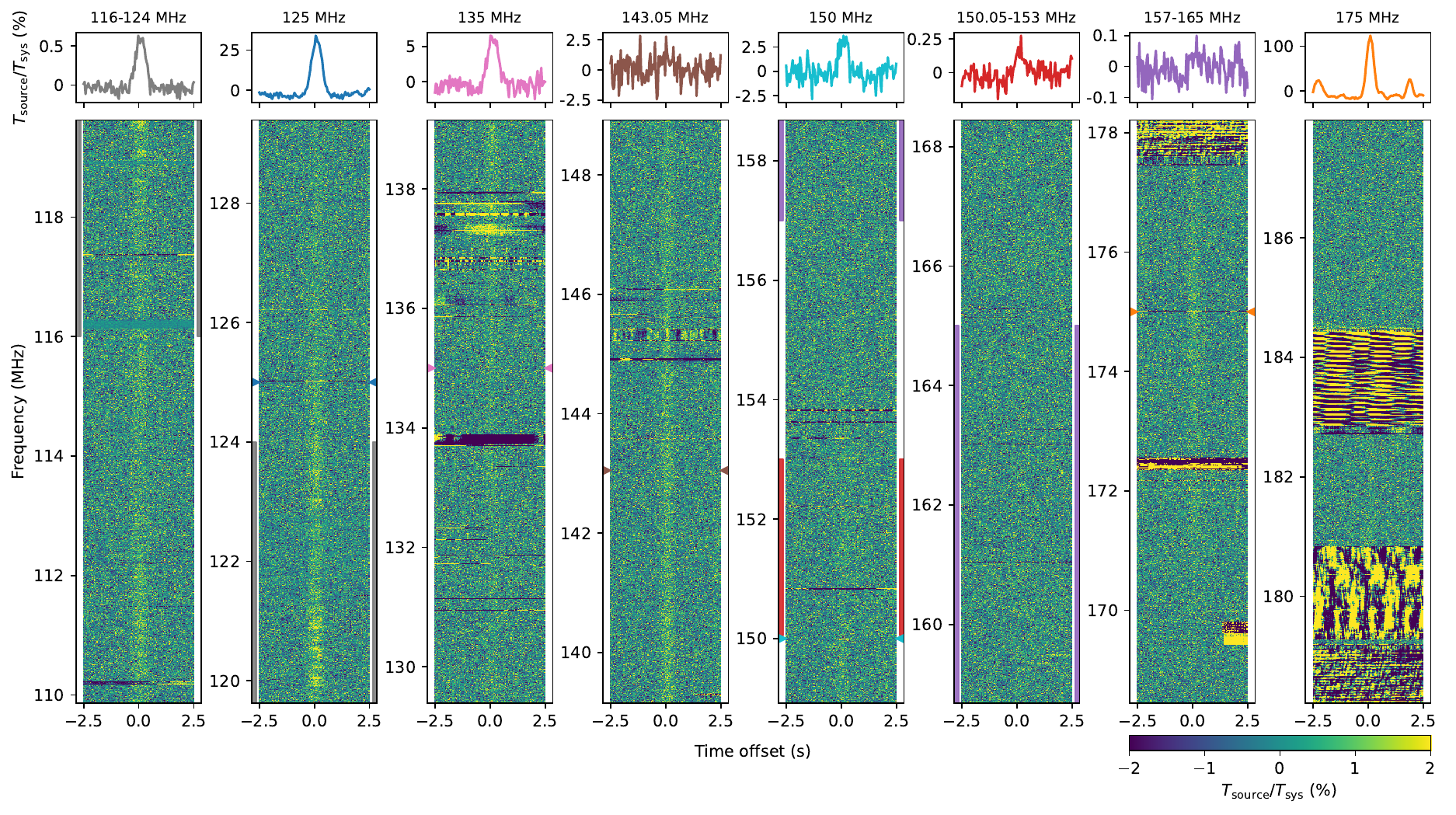}
  \includegraphics[width=\textwidth]{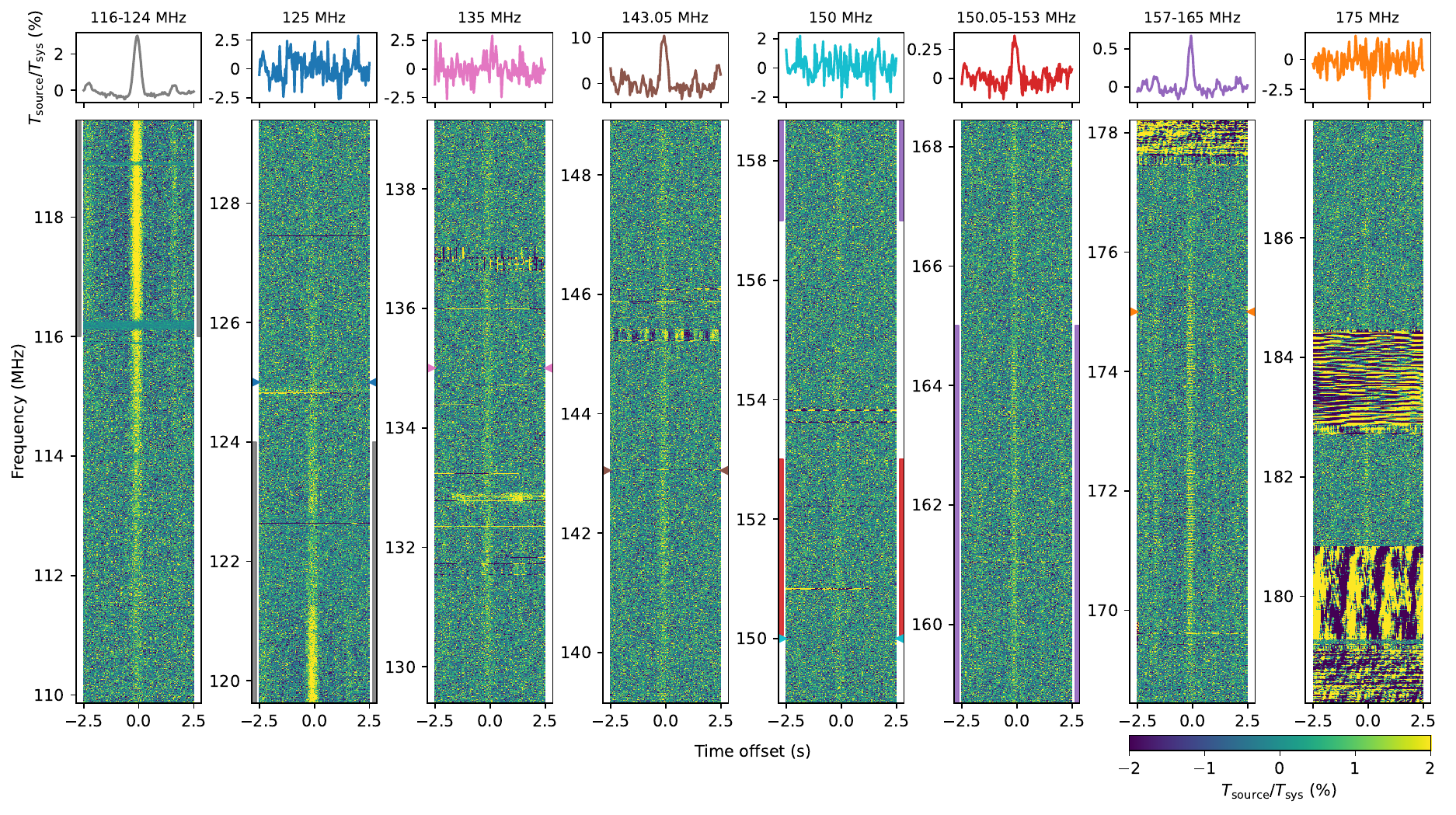}
  \caption{Spectral and temporal properties of the passes of satellites 45186 (average of 11 TABs) and 51998 (average of 10 TABs). For each satellite pass, normalised, aligned and averaged dynamic spectra are shown over the entire observed bandwidth and within 2.5~s on the predicted passage time. Time series at narrow-band frequencies of 125, 135, 143.05, 150, and 175~MHz are shown in the top insets, as well as for broad-band frequency ranges (116 to 124, 150.05 to 153, and 157 to 165~MHz). The colour of each time series matches the marked frequencies and frequency ranges in the same colours to the sides of the dynamic spectra. For both satellites a combination of broad-band and narrow-band emission is visible. In the case of satellite 45186, broad-band emission is mostly confined to the frequencies below 155~MHz, but narrow-band emission at 125, 135, 150, and 175~MHz is detected, with sidelobes being visible at 175~MHz. Some structure in the broad-band emission is obvious between 120 and 122~MHz. For satellite 51998, broad-band emission is clear at all frequencies not affected by terrestrial signals, while narrow-band emission is absent, except for 143.05~MHz. Between 170 and 176~MHz, a comb of narrow-band, regularly spaced peaks, is superposed on the broad-band emission. The temporal profiles show time offsets of the observed satellite pass with respect to predictions ($+0.09$~s for 45186, $-0.07$~s for 51998).}
  \label{fig:zoomed_plots}
\end{figure*}

Inspection of these averages of TABs showed broad-band emission throughout the observed frequency range, coinciding with the crossing times of Starlink satellites. Normalised, aligned, and averaged dynamic spectra for two satellites are shown in Fig.~\ref{fig:zoomed_plots}. The dynamic spectra have a time resolution of 41~ms and the full frequency resolution of 12.21~kHz. Due to the normalisation with the dynamic spectra of the other TABs, bright signals in those TABs may lead to depressions in these plots. To prevent masking of signals associated with satellites, no masking has been applied when normalising, aligning and averaging these spectra. Not all satellites reveal broad-band emission at the same frequencies -- the two most common frequency ranges where emission is detected are at 116 to 124~MHz and 157 to 165~MHz. We focus our analysis on these two frequency ranges, but also include the ITU-R RAS frequency band from 150.05 to 153~MHz.

Besides broad-band emission, narrow-band emission was also detected in, and confined to, several individual 12.21~kHz channels. The frequencies of these channels cover 124.994 to 125.006~MHz, 134.991 to 135.004~MHz, 143.048 to 143.060~MHz, 149.994 to 150.006~MHz and 174.994 to 175.006~MHz. We include these signals in our analysis, and will refer to them as the narrow-band emission at 125, 135, 143.05, 150 and 175~MHz. As the maximum radial velocities of the Starlink satellites in this observation are less than $|v_\mathrm{r}|<1$~km\,s$^{-1}$, any Doppler shifts at these frequencies are less than $\sim600$~Hz and hence confined to individual spectral channels.

As shown in Fig.~\ref{fig:zoomed_plots}, the signal strength of these narrow-band emission can vary significantly between frequencies as well as satellites. In some cases, the narrow-band features were so bright, that the satellite was detected passing through the sidelobes of individual TABs. Furthermore, in many cases, especially at 125~MHz, the narrow-band signals were superposed with terrestrial signals. This is also why the data processing strategy had to be modified in order to extract the narrow-band signals properly. Instead of subtracting the average of all beams from each spectrogram we subtracted a spectral baseline in a small window around each narrow-band peak. 

Finally, in some, but not all, of the lower altitude Starlink satellites, a comb of narrow (within a 12.21~kHz channel) peaks was seen in the frequency range above 155~MHz. The dynamic spectra of satellite 51998 shown in Fig.~\ref{fig:zoomed_plots} shows this comb for frequencies between 170 and 176~MHz. Power spectra of the emission between 157 to 165~MHz shows that these peaks are spaced at 50~kHz offsets and is detectable in 17 of the 46 satellites at lower altitudes, but none of the higher altitude satellites. The satellites where this comb was detected are marked in Table~\ref{tab:satellite_passes}.

For all satellites which were detected through either broad-band or narrow-band emission, we determined the time offset between the observed and the predicted passage time through the TABs by fitting a Gaussian profile to the temporal emission profiles. These time offsets are listed in Table~\ref{tab:satellite_passes}. We found that the time offsets are less than 1~s for all but four satellites, and excluding those yields a median time offset of $\bar{\Delta t}=-0.03\pm0.14$~s. The four satellites with the largest time offsets passed through the beam pattern by as much as 6.4~s earlier, and others 1.3~s late compared to predictions. We furthermore found that the temporal width of the Gaussian fits matches those from predictions, where the satellites at 350~km orbital altitudes moved through the beam faster than those at 550~km. Subsequently, all these offsets were used to modify the satellite ephemerides and further analyses were based on the corrected positions.

\begin{figure}[!tp]
    \centering
    \includegraphics[width=\columnwidth, viewport=40 40 650 650, clip=]{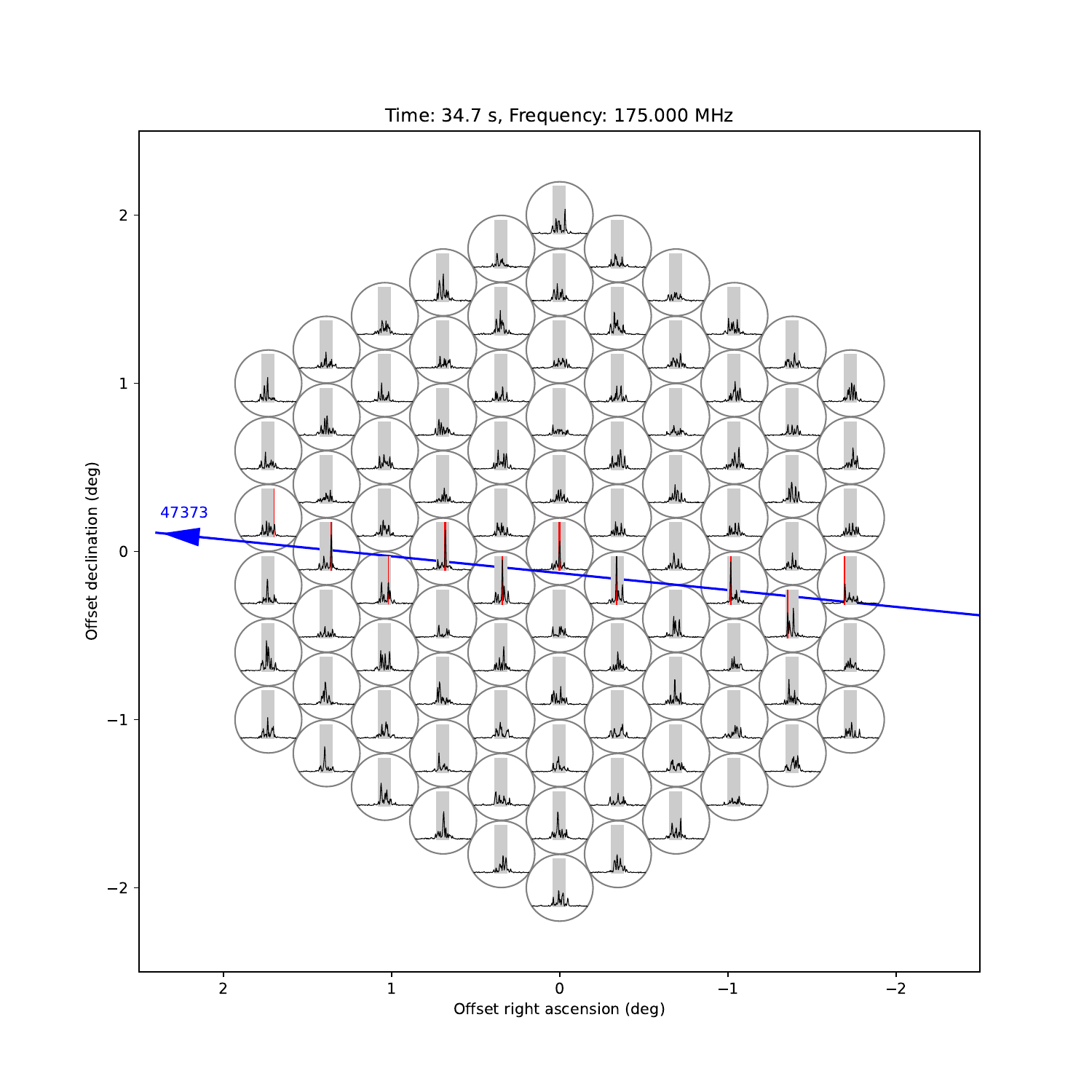}
    \caption{Visualisation of the detected signal from satellite 47373 as it crossed the field of view of the LOFAR tied array beam pattern. Each circle marks one of the beams. The inlays show the 175~MHz signal (spectral PFD) as a function of time spanning about $\sim$20~seconds centred around the event time (approximately 35~s after observation start). The grey-shaded areas mark the total time interval over which the satellite was in the field of view, the red shaded areas refer to the time when the satellite was in the corresponding beam area.}
    \label{fig:beam_plot_spectra_47373}
\end{figure}

\begin{figure}[!tp]
    \centering
    \includegraphics[width=\columnwidth, viewport=40 40 650 650, clip=]{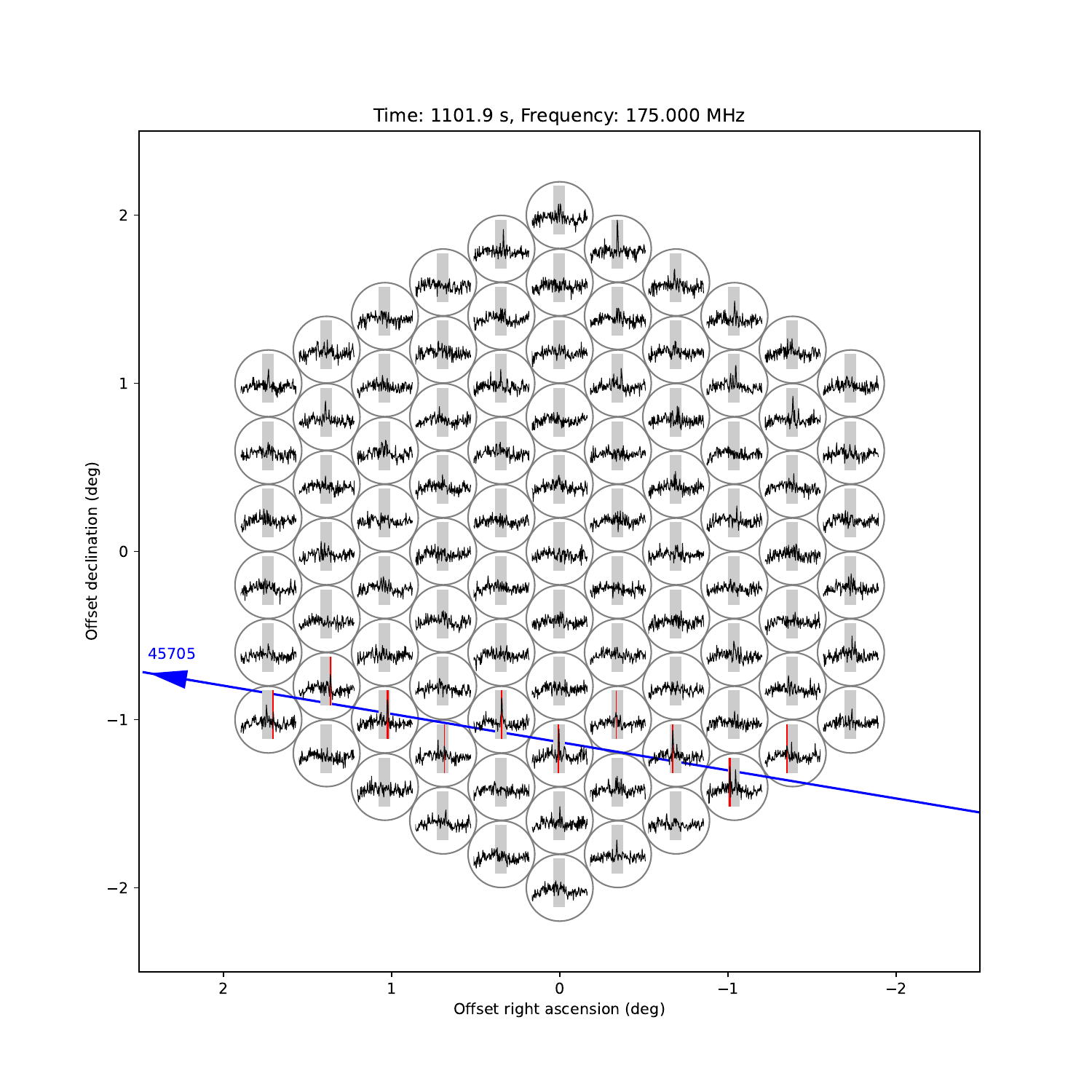}
    \caption{As Fig.~\ref{fig:beam_plot_spectra_47373} but for satellite 45705.}
    \label{fig:beam_plot_spectra_45705}
\end{figure}

\begin{figure}[!tp]
    \centering
    \includegraphics[width=\columnwidth, viewport=40 40 650 650, clip=]{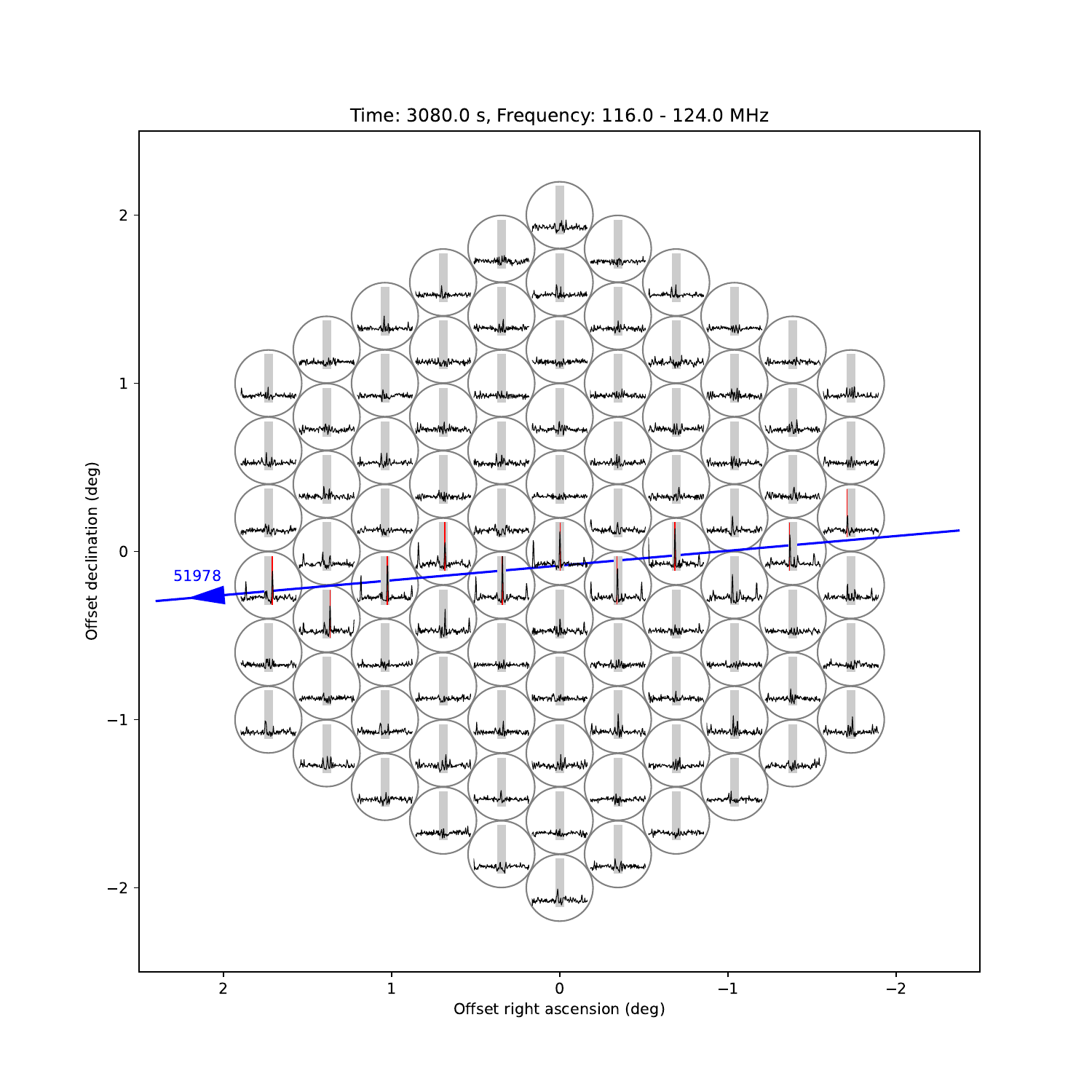}
    \caption{As Fig.~\ref{fig:beam_plot_spectra_47373} but for satellite 51978 visualising a broad-band signal in 116$-$124~MHz.}
    \label{fig:beam_plot_spectra_51978}
\end{figure}

To visualise the emission as a satellite passes through the LOFAR TAB beam pattern, Figs.~\ref{fig:beam_plot_spectra_47373}, \ref{fig:beam_plot_spectra_45705} and \ref{fig:beam_plot_spectra_51978} show the temporal profiles of satellite passes in comparison to the location of the satellite as it passes through the beam pattern. The case of satellite 47373 shown in Fig.~\ref{fig:beam_plot_spectra_47373} is one of example for a very bright event, where the narrow-band emission at 175~MHz was strong enough to be detected in all TABs for the full duration that the satellite passed through the $4\fdg7$ FWHM station beam. In other cases, such as for the pass of satellite 45705 (Fig.~\ref{fig:beam_plot_spectra_45705}) the behaviour was `normal', however, and a signal was only detected in the beams covering the satellite sky track. For completeness, also an example for the broad-band emission between 116 to 124~MHz is displayed in Fig.~\ref{fig:beam_plot_spectra_51978} for the pass of satellites 51978. As expected, the strongest detections coincide with the predicted time that the satellites passed through the individual TABs, confirming that the signal was coming from the direction of the satellites.

\begin{figure*}[!tp]
    \includegraphics[width=\columnwidth, viewport=10 25 420 540, clip]{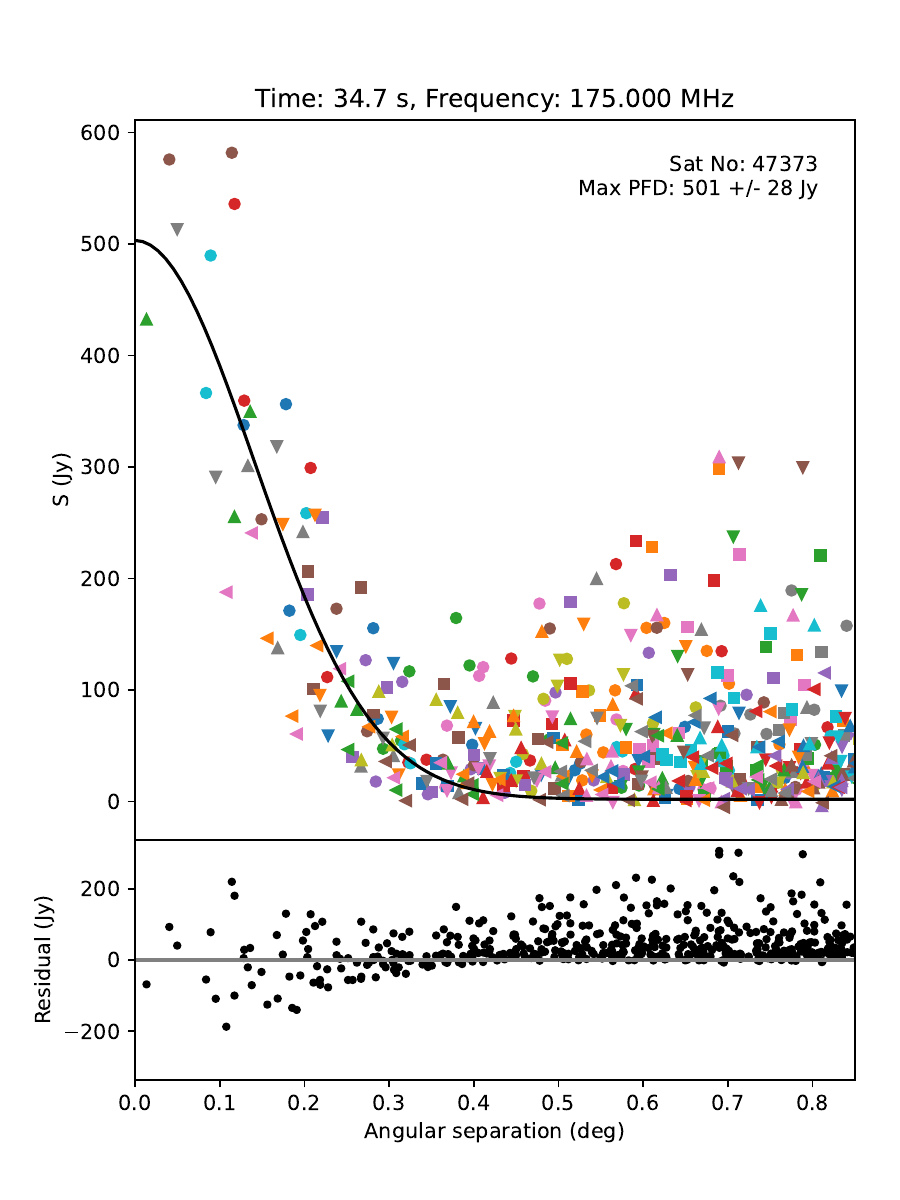}
    \includegraphics[width=\columnwidth, viewport=10 25 420 540, clip]{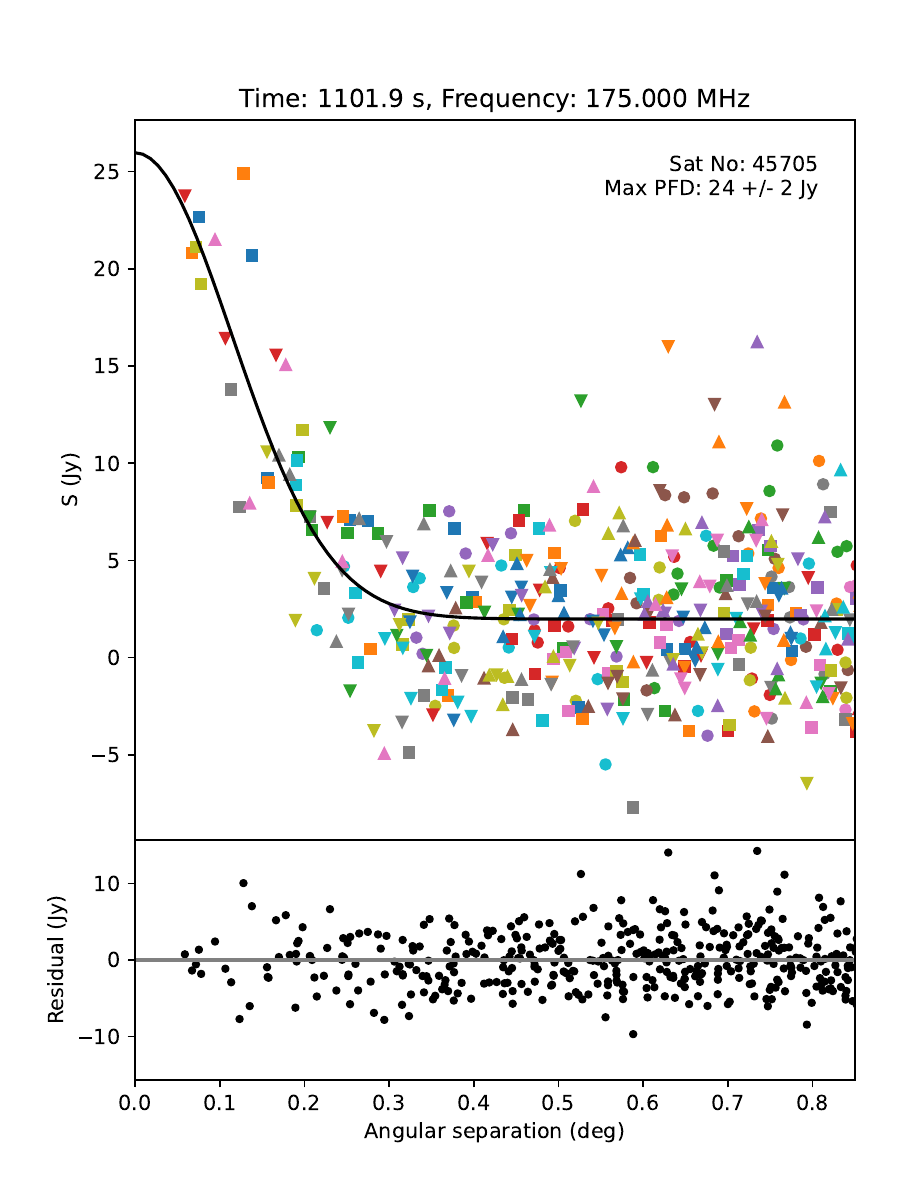}
    \caption{Measured spectral power flux densities for satellites 47373 and 45705 as a function of angular separation from beam centres. Different colours and symbols mark different beams. The black solid line represents a least-squares fit (Gaussian function) to the data points. This allows to estimate the actual spectral PFD of the satellite, which is about 460~Jy (47373) and 25~Jy (45705), respectively, averaged over one spectral channel of 12.2~kHz at 175~MHz.}
    \label{fig:beam_plot_example}
\end{figure*}

\begin{figure*}[!tp]
    \includegraphics[width=\textwidth]{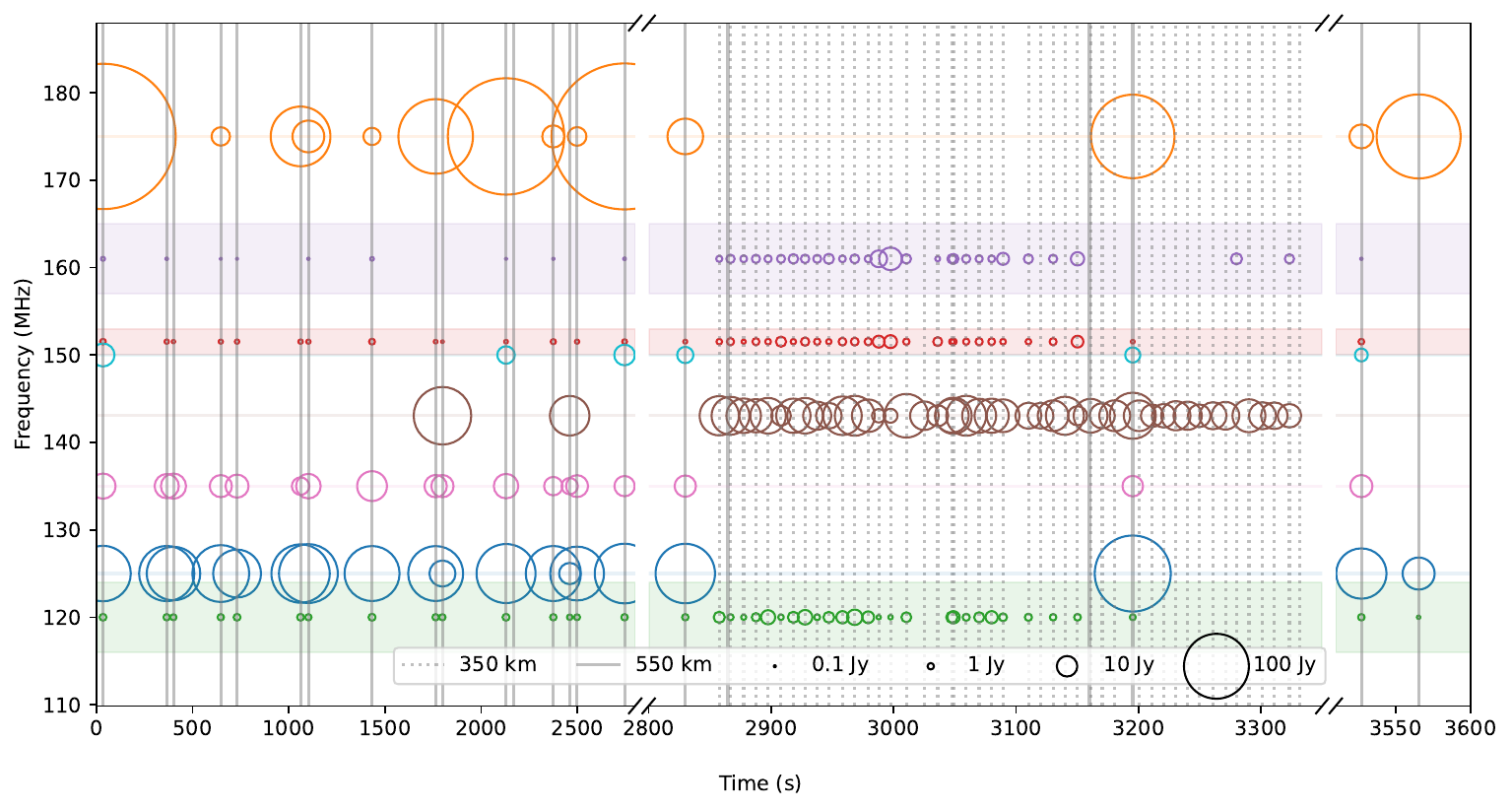}
    \caption{Radio emission detected from Starlink satellites during the LOFAR observation. Passes of Starlink satellites through the LOFAR beam pattern are marked in time with solid vertical lines for satellites at $h=550$~km, and dotted lines for those at $h=350$~km. The coloured horizontal lines and bands indicate the frequencies of frequency ranges in which fluxes were measured, with the circles indicating the corresponding flux density measurements.}
    \label{fig:signal_detections}
\end{figure*}

Next, we used the intensity-calibrated spectra to estimate the power flux densities (PFD) for each one of the detected signals. As the satellites usually did not cross any of the beam centres exactly, we determined the PFD as a function of the angular separation between the satellite positions with respect to each of the TAB centres; see Figure~\ref{fig:beam_plot_example} for two example satellites. 
Based on a Gaussian least-squares fit to the data points, the peak PFD could be estimated. These PFD measurements are provided in Table~\ref{tab:satellite_passes}. Furthermore, a visual overview is provided in Fig.~\ref{fig:signal_detections}. It is noteworthy that for the events with very high intensity (above about 100~Jy) the Gaussian fit was made difficult because of the cross-talk induced by the LOFAR beam-former (e.g. Fig.~\ref{fig:beam_plot_example} left panel). Therefore, the width parameter of the Gaussian fit curve was constrained to values below $1.1\vartheta_\mathrm{fwhm}^\mathrm{tab}$. Likewise, for all fits the zero level offset was constrained to values close to zero. Also, the scatter in the flux density values was rather large, such that the accuracy of the $S_\nu$ values in Table~\ref{tab:satellite_passes} should not be overestimated.

\begin{table*}{}
\tiny
\centering
\caption{Properties of the Starlink satellite passes through the LOFAR beam pattern. Satellites are identified by their NORAD catalogue number and COSPAR-based international designator (which provides the launch year, launch number and sequential identifier). For each satellite, the distance $d$ between the satellite and LOFAR is given at the mid-point of the pass through the $4\fdg7$ FHWM station beam. This midpoint is given by $t_\mathrm{mid}$, from the start of the observation at 18:30:00\,UTC on 2022 April 1. The duration of the passage of the station beam is given by $t_\mathrm{pass}$, while $\Delta t$ represents the time offset between the predicted and measured time of the pass. The $n_\mathrm{TAB}$ column indicates the number of tied-array beams the satellite passed through. Detected flux densities $S_\nu$ are provided for different frequencies or frequency ranges over the LOFAR observing band. The $\dagger$ symbol indicates that a comb with 50~kHz spacing was detected in the 157 to 165~MHz frequency range.}
\label{tab:satellite_passes}
\begin{tabular}{l|rrrrr|cccccccc}
\hline
NORAD / COSPAR & $d$ & $t_\mathrm{mid}$ & $t_\mathrm{pass}$ & 
$\Delta t$ (s) & $n_\mathrm{TAB}$ & \multicolumn{6}{c}{$S_\nu$ (Jy)} \\
& (km) & (s) & (s) & (s) & & 116--124 & 125 & 135 & 143 & 150 & 150.05--153 & 157--165 & 175 \\
& & & & & & (MHz) & (MHz) & (MHz) & (MHz) & (MHz) & (MHz) & (MHz) & (MHz) \\
\hline
47373 / 2021$-$005AA &  554.7 &    34.66 &  6.26 & $+0.00$ &  11 & 1.1(1) & 73(3) & 15(1) &            & 13(2) & 0.8(1) & 0.4(1)          & 501(28) \\
47371 / 2021$-$005Y  &  553.6 &   365.68 &  5.04 & $+0.05$ &   8 & 1.0(1) & 72(3) & 14(1) &            &            & 0.5(1) & 0.2(1)          &            \\
47597 / 2021$-$009BB &  553.7 &   400.77 &  3.83 & $-0.01$ &   4 & 0.9(1) & 68(5) & 15(2) &            &            & 0.3(1) &                       &            \\
47645 / 2021$-$012AB &  553.6 &   646.62 &  5.83 & $+0.15$ &  13 & 1.0(1) & 77(3) & 12(1) &            &            & 0.4(1) & 0.1(1)          & 8(1) \\
47595 / 2021$-$009AZ &  553.8 &   731.25 &  6.13 & $+0.14$ &  11 & 1.1(1) & 54(3) & 13(1) &            &            & 0.5(1) & 0.1(1)          &            \\
47596 / 2021$-$009BA &  554.5 &  1061.19 &  5.56 & $+0.04$ &  10 & 1.1(1) & 81(4) & 7(1) &            &            & 0.5(1) &                       & 85(6) \\
45705 / 2020$-$035BA &  554.4 &  1101.86 &  5.50 & $-0.06$ &  10 & 1.0(1) & 82(3) & 14(1) &            &            & 0.5(1) & 0.1(1)          & 24(2) \\
45661 / 2020$-$035E  &  553.4 &  1432.54 &  6.11 & $-0.02$ &  11 & 1.2(1) & 72(3) & 21(1) &            &            & 0.8(1) & 0.3(1)          & 7(1) \\
45677 / 2020$-$035W  &  553.1 &  1763.14 &  3.76 & $-0.08$ &   4 & 1.0(1) & 72(5) & 12(2) &            &            & 0.3(2) &                       & 132(13) \\
45235 / 2020$-$012BK &  553.3 &  1798.44 &  4.88 & $-0.01$ &   7 & 0.9(1) & 16(1) & 12(1) & 79(5) &            & 0.2(1) &                       &            \\
45186 / 2020$-$012J  &  554.0 &  2128.86 &  6.26 & $+0.09$ &  11 & 1.2(1) & 83(3) & 14(1) &            & 7(1) & 0.3(1) & 0.1(1)          & 322(18) \\
47621 / 2021$-$012B  &  555.4 &  2167.83 &  1.07 &         &   0 &              &            &            &            &            &              &                       &            \\
47666 / 2021$-$012AY &  555.0 &  2374.54 &  5.36 & $+0.11$ &   9 & 1.0(1) & 72(3) & 8(1) &            &            & 0.6(1) & 0.1(1)          & 12(2) \\
45187 / 2020$-$012K  &  555.6 &  2459.64 &  4.40 & $+0.07$ &   6 & 0.7(1) & 11(2) & 6(2) & 37(5) &            &              &                       &            \\
47625 / 2021$-$012F  &  554.6 &  2498.09 &  6.02 & $+0.65$ &  12 & 0.8(1) & 69(3) & 11(1) &            &            & 0.4(1) &                       & 8(1) \\
47651 / 2021$-$012AH &  554.4 &  2744.72 &  6.15 & $+0.02$ &  11 & 1.0(1) & 84(4) & 10(1) &            & 10(1) & 0.6(1) & 0.2(1)          & 506(29) \\
47620 / 2021$-$012A  &  554.4 &  2829.89 &  5.88 & $-0.05$ &  10 & 1.0(1) & 84(3) & 11(1) &            & 6(2) & 0.4(1) &                       & 30(2) \\
52003 / 2022$-$025AZ &  356.1 &  2857.64 &  2.98 & $+0.07$ &   6 & 3.0(2) &            &            & 37(4) &            & 0.8(2) & 0.9(1)          &            \\
47636 / 2021$-$012S  &  553.8 &  2864.70 &  1.77 &         &   0 &              &            &            &            &            &              &                       &            \\
52002 / 2022$-$025AY &  356.2 &  2866.81 &  3.11 & $-0.03$ &   8 & 0.9(1) &            &            & 34(4) &            & 1.2(2) & 1.7(2)          &            \\
52000 / 2022$-$025AW &  356.5 &  2877.57 &  3.21 & $-0.09$ &   8 & 0.6(1) &            &            & 28(3) &            & 0.4(1) & 1.1(1)          &            \\
52001 / 2022$-$025AX &  356.1 &  2887.59 &  3.32 & $-0.12$ &   9 & 1.7(1) &            &            & 26(3) &            & 1.2(2) & 1.6(1)$\dagger$ &            \\
51998 / 2022$-$025AU &  356.4 &  2897.50 &  3.38 & $-0.07$ &  10 & 4.8(2) &            &            & 30(3) &            & 0.8(1) & 1.3(1)$\dagger$ &            \\
51999 / 2022$-$025AV &  356.6 &  2908.06 &  3.47 & $+0.01$ &  10 & 0.8(1) &            &            & 9(2) &            & 2.5(2) & 1.6(1)$\dagger$ &            \\
51996 / 2022$-$025AS &  356.5 &  2918.15 &  3.54 & $-0.06$ &  11 & 3.0(1) &            &            & 28(3) &            & 0.7(1) & 2.2(1)          &            \\
51994 / 2022$-$025AQ &  356.7 &  2927.70 &  3.60 & $-0.02$ &  11 & 5.1(2) &            &            & 30(3) &            & 1.7(2) & 1.8(1)$\dagger$ &            \\
51992 / 2022$-$025AN &  356.7 &  2937.59 &  3.67 & $-0.06$ &  11 & 0.8(1) &            &            & 19(2) &            & 1.0(1) & 1.6(1)$\dagger$ &            \\
51997 / 2022$-$025AT &  356.3 &  2947.11 &  3.72 & $-0.11$ &  11 & 2.8(1) &            &            & 16(2) &            & 0.7(1) & 2.5(2)$\dagger$ &            \\
51995 / 2022$-$025AR &  356.4 &  2958.22 &  3.76 & $-0.02$ &  11 & 3.6(1) &            &            & 35(3) &            & 1.4(1) & 1.1(1)$\dagger$ &            \\
51990 / 2022$-$025AL &  356.4 &  2968.25 &  3.81 & $-0.03$ &  11 & 5.7(2) &            &            & 38(4) &            & 1.5(1) & 1.9(1)          &            \\
51993 / 2022$-$025AP &  356.8 &  2979.25 &  3.84 & $-0.04$ &  11 & 3.3(1) &            &            & 25(3) &            & 1.3(1) & 1.2(1)$\dagger$ &            \\
51988 / 2022$-$025AJ &  356.6 &  2987.83 &  3.89 & $+0.33$ &  11 & 0.5(1) &            &            & 4(2) &            & 3.4(2) & 6.7(4)$\dagger$ &            \\
51986 / 2022$-$025AG &  356.6 &  2997.48 &  3.92 & $+0.04$ &  11 & 0.5(1) &            &            & 5(2) &            & 4.0(3) & 12.3(7)          &            \\
51991 / 2022$-$025AM &  356.6 &  3010.40 &  3.94 & $-1.73$ &  11 & 2.4(1) &            &            & 45(3) &            & 1.0(1) & 2.1(1)$\dagger$ &            \\
51984 / 2022$-$025AE &  357.1 &  3025.10 &  3.97 & $-6.38$ &  11 &              &            &            & 19(2) &            &              &                       &            \\
51989 / 2022$-$025AK &  357.5 &  3035.99 &  4.00 & $-2.99$ &  11 &              &            &            & 11(2) &            & 1.9(2) & 0.6(1)          &            \\
51987 / 2022$-$025AH &  357.9 &  3047.57 &  4.02 & $+1.29$ &  11 & 3.1(2) &            &            & 27(3) &            & 0.8(1) & 2.0(2)$\dagger$ &            \\
51982 / 2022$-$025AC &  356.9 &  3048.97 &  4.01 & $-0.12$ &  11 & 3.6(1) &            &            & 31(3) &            & 0.9(1) & 2.3(2)$\dagger$ &            \\
51980 / 2022$-$025AA &  356.7 &  3059.27 &  4.01 & $-0.05$ &  11 & 1.2(1) &            &            & 38(4) &            & 0.9(1) & 1.6(1)$\dagger$ &            \\
51985 / 2022$-$025AF &  357.1 &  3069.85 &  4.02 & $-0.01$ &  11 & 2.4(1) &            &            & 27(3) &            & 1.2(1) & 1.3(1)          &            \\
51978 / 2022$-$025Y  &  357.1 &  3080.04 &  4.02 & $-0.00$ &  11 & 3.6(2) &            &            & 28(2) &            & 1.1(1) & 1.1(1)$\dagger$ &            \\
51983 / 2022$-$025AD &  356.9 &  3089.36 &  4.00 & $-0.05$ &  11 & 1.5(1) &            &            & 25(2) &            & 0.8(1) & 3.5(2)$\dagger$ &            \\
51981 / 2022$-$025AB &  356.9 &  3110.07 &  3.98 & $-0.03$ &  11 & 1.2(1) &            &            & 16(2) &            & 0.8(1) & 2.0(1)$\dagger$ &            \\
51975 / 2022$-$025V  &  357.1 &  3119.98 &  3.96 & $-0.12$ &  11 &              &            &            & 16(2) &            &              &                       &            \\
51976 / 2022$-$025W  &  357.4 &  3130.23 &  3.93 & $-0.06$ &  11 & 1.0(1) &            &            & 22(3) &            & 1.2(1) & 1.6(1)$\dagger$ &            \\
51972 / 2022$-$025S  &  357.5 &  3139.86 &  3.91 & $-0.05$ &  11 &              &            &            & 34(4) &            &              &                       &            \\
51979 / 2022$-$025Z  &  357.4 &  3150.16 &  3.87 & $-0.09$ &  11 & 1.1(1) &            &            & 9(2) &            & 3.3(2) & 4.2(3)          &            \\
47677 / 2021$-$012BK &  554.7 &  3159.90 &  2.76 &         &   1 &              &            &            &            &            &              &                       &            \\
51970 / 2022$-$025Q  &  357.4 &  3160.85 &  3.82 & $-0.08$ &  11 &              &            &            & 27(3) &            &              &                       &            \\
51973 / 2022$-$025T  &  357.6 &  3170.33 &  3.78 & $-0.12$ &  11 &              &            &            & 15(2) &            &              &                       &            \\
51968 / 2022$-$025N  &  357.4 &  3180.57 &  3.73 & $-0.12$ &  12 &              &            &            & 23(2) &            &              &                       &            \\
47661 / 2021$-$012AT &  555.3 &  3195.27 &  5.88 & $-0.02$ &  11 & 0.8(1) & 137(60) & 10(1) & 50(4) & 5(1) & 0.3(1) &                       & 166(10) \\
51966 / 2022$-$025L  &  357.8 &  3200.51 &  3.60 & $-0.08$ &  10 &              &            &            & 22(3) &            &              &                       &            \\
51971 / 2022$-$025R  &  357.5 &  3211.10 &  3.52 & $-0.06$ &  10 &              &            &            & 12(2) &            &              &                       &            \\
51964 / 2022$-$025J  &  357.7 &  3220.55 &  3.45 & $-0.04$ &  10 &              &            &            & 13(2) &            &              &                       &            \\
51969 / 2022$-$025P  &  357.7 &  3230.47 &  3.36 & $-0.08$ &   9 &              &            &            & 20(3) &            &              &                       &            \\
51967 / 2022$-$025M  &  357.5 &  3240.09 &  3.26 & $-0.08$ &   8 &              &            &            & 19(3) &            &              &                       &            \\
51962 / 2022$-$025G  &  357.7 &  3249.53 &  3.14 & $+0.09$ &   8 &              &            &            & 12(3) &            &              &                       &            \\
51965 / 2022$-$025K  &  358.1 &  3260.78 &  3.04 & $-0.06$ &   6 &              &            &            & 18(2) &            &              &                       &            \\
51960 / 2022$-$025E  &  358.0 &  3270.97 &  2.92 & $-0.08$ &   6 &              &            &            & 18(3) &            &              &                       &            \\
51957 / 2022$-$025B  &  358.0 &  3279.97 &  2.76 & $-0.03$ &   5 &              &            &            &            &            &              & 3.0(3)          &            \\
51959 / 2022$-$025D  &  357.8 &  3290.15 &  2.59 & $+0.65$ &   4 &              &            &            & 26(4) &            &              &                       &            \\
51963 / 2022$-$025H  &  358.2 &  3301.21 &  2.42 & $-0.10$ &   3 &              &            &            & 18(5) &            &              &                       &            \\
51961 / 2022$-$025F  &  358.0 &  3310.88 &  2.21 & $-0.02$ &   2 &              &            &            & 17(5) &            &              &                       &            \\
51958 / 2022$-$025C  &  358.3 &  3323.21 &  1.95 & $-0.17$ &   1 &              &            &            & 13(4) &            &              & 2.1(3)          &            \\
51956 / 2022$-$025A  &  358.6 &  3331.34 &  1.75 &         &   1 &              &            &            &            &            &              &                       &            \\
47670 / 2021$-$012BC &  557.0 &  3527.13 &  5.93 & $+0.05$ &  11 & 1.0(1) & 60(5) & 12(1) &            & 4(1) & 0.8(1) & 0.1(1)          & 14(2) \\
45583 / 2020$-$025BE &  557.0 &  3565.40 &  3.26 & $-0.03$ &   2 & 0.3(1) & 24(6) &            &            &            &              &                       & 168(28) \\
\hline
\end{tabular}
\end{table*}

\section{Analysis of the detected events}\label{sec:analysis}

\subsection{Signal properties}\label{ssec:signal_properties}
Using the flux density measurements of satellite events at the narrow- and broad-band frequencies as listed in Table~\ref{tab:satellite_passes} we can infer some properties of the detected signals.

We found that the narrow-band emission at 125, 135, 150, and 175~MHz is only detected for the Starlink satellites at their operational altitude of $h=550$~km, and not seen in any of the Starlink satellites in the lower orbit at altitudes of $h=350$~km. As the higher altitude satellites are more distant ($d\sim555$~km) compared to the lower altitude satellites ($d\sim356$~km), any emission of equal intensity should cause a detection in the received data about $(555~\mathrm{km}/356~\mathrm{km})^2\sim2.4$ times brighter for the satellites at lower altitudes. While the individual satellites showed some variation in the signal strengths, it is deemed extremely unlikely that all of the lower altitude satellites would by chance have very low emission. Hence, it naturally appears that there is an intrinsic difference between the satellites in higher altitude and lower altitude orbits with respect to the narrow-band features.

This is not the case for the broad-band emission, which was detected for the majority of satellites, regardless of their orbital altitude. We found that the median PFD of the low altitude satellites is a factor 2.0 and 2.3 higher than that of the high altitude satellites for frequency ranges of 116 to 124~MHz and 150.05 to 153~MHz, respectively. As this is close to the expected factor of 2.4, this indicates that the generation of this emission is independent on altitude. Curiously, the broad-band emission between 157 and 165~MHz is a factor 15 higher in the low altitude satellites, suggesting an intrinsic difference in this frequency range.

The occurrence of the signals for individual satellites at different frequencies is correlated. For 18 out of 19 cases in which narrow-band emission at 125~MHz was detected, emission was also present at 135~MHz, albeit somewhat fainter. A similar relation exists between the emission at 125~MHz and 175~MHz, though the emission at 175~MHz appears to be more variable and can be brighter than at 125~MHz. The signal at 175~MHz was detected in 14 cases. The narrow-band emission at 150~MHz was only seen for those satellites that were very bright at 175~MHz (and cross the station beam) and was detected in six cases.

As the lower altitude satellites were still in the orbit-raising phase, the 125~and 175~MHz might be associated with the regular operation (e.g.\ communication-link transmissions) of the satellites. Also, both frequencies are odd multiples of 25~MHz -- a frequency often used for local oscillators -- and could be harmonics, which usually appear stronger at either odd or even multiples of the fundamental mode. This would also explain, why the 150~MHz signal is only present for the brightest of the 175~MHz detections (as 150~MHz is an even multiple of 25~MHz). Typically, square wave-like signals are expected to produce odd harmonics. It is unclear, how the 135~MHz feature would fit into this. It might be owing to some intermodulation product of the detected narrow-band features with some other signal, but we were not able to find further evidence for this.

We attribute the narrow-band emission detected at a frequency of 143.05~MHz to the GRAVES space surveillance radar \citep{michal+05}. The GRAVES transmitter is located 30~km east of Dijon, France and is known to transmit continuous wave signals at 143.050~MHz for bi-static Doppler tracking of satellites. The transmitter illuminates a $180\degr$ range in azimuth (east to west through south) and a $30\degr$ range in elevation \citep{jouade_barka19}. Though the radiated power of the transmitter is not publicly known, radar reflections from meteors are regularly detected by radio amateurs using modest equipment, even for meteors located well outside of the nominal illumination pattern of the GRAVES transmitter \citep[e.g.][]{fleet15}. The Starlink satellites that we observed were also located far outside of the (known) GRAVES illumination area, implying that even in the far sidelobes of the GRAVES radar the effectively transmitted power is substantial.

Another interesting finding is that most high-altitude satellites do not show GRAVES reflections, even though LOFAR should have the sensitivity to detect them. The two satellites that were detected at 143.05~MHz were even brighter than the low-altitude satellite reflections (when they should be weaker owing to longer propagation paths). This suggests that the details of the propagation are subject to several effects, the magnitude of which cannot easily be determined without additional information. One aspect is certainly the orientation of the satellite relative to the LOFAR station. It is known that Starlink uses the `open-book' mode during orbit raising, where the solar array is aligned parallel to the satellite body to reduce atmospheric drag, while the operational satellites are in `shark-fin' configuration, where the solar array is located mostly behind the satellite as seen from Earth. Furthermore, the exact path geometry is expected to differ between lower and higher orbit altitudes, as well as the side-lobe gain of GRAVES towards different elevations.

\subsection{Assessment of transmitted power levels}

\begin{table*}[!tp]
\caption{Derived satellite transmitter parameters for the weakest and brightest detections.}
\label{tab:sat_tx_parameters}
\centering
\tiny
\begin{tabular}{l r r c c c c c c}
\hline\hline
\rule{0ex}{3ex}Frequency & Type & Altitude & \multicolumn{2}{c}{$P_\nu^\mathrm{tx,eirp}$} & \multicolumn{4}{c}{E-field$^\mathrm{(1)}$}  \\
(MHz) & & (km) & \multicolumn{2}{c}{$\left(\mathrm{dB}\left[\mathrm{W\,Hz}^{-1}\right]\right)$} & \multicolumn{4}{c}{$\left(\mathrm{dB}\left[\mu\mathrm{V\,m}^{-1}\right]\right)$}\\[1ex]\hline
\rule{0ex}{3ex}& & & & & \multicolumn{2}{c}{$\Delta f_\mathrm{det}=120~\mathrm{kHz}$} & \multicolumn{2}{c}{$\Delta f_\mathrm{det}=2.95~\mathrm{MHz}$}\\
& & & min & max & min & max & min & max\\[1ex]
\hline
\rule{0ex}{3ex}116--124    & broad-band  & $\sim$360 & $-$141 & $-$130 & 24 & 35 & 38 & 49 \\ 
                           & broad-band  & $\sim$550 & $-$139 & $-$133 & 26 & 32 & 40 & 46 \\\hline 
\rule{0ex}{3ex}125         & narrow-band & $\sim$360 &        &        &    &    &    &    \\ 
                           & narrow-band & $\sim$550 & $-$124 & $-$113 & 32 & 43 & 32 & 43 \\\hline  
\rule{0ex}{3ex}135         & narrow-band & $\sim$360 &        &        &    &    &    &    \\ 
                           & narrow-band & $\sim$550 & $-$126 & $-$121 & 30 & 35 & 30 & 35 \\\hline  
\rule{0ex}{3ex}143         & narrow-band & $\sim$360 & $-$132 & $-$121 & 24 & 34 & 24 & 34 \\ 
                           & narrow-band & $\sim$550 & $-$118 & $-$115 & 37 & 40 & 37 & 40 \\\hline  
\rule{0ex}{3ex}150         & narrow-band & $\sim$360 &        &        &    &    &    &    \\ 
                           & narrow-band & $\sim$550 & $-$128 & $-$123 & 28 & 33 & 28 & 33 \\\hline  
\rule{0ex}{3ex}150.05--153 & broad-band  & $\sim$360 & $-$142 & $-$132 & 24 & 34 & 38 & 48 \\ 
                           & broad-band  & $\sim$550 & $-$141 & $-$135 & 24 & 31 & 38 & 45 \\\hline  
\rule{0ex}{3ex}157--165    & broad-band  & $\sim$360 & $-$140 & $-$127 & 25 & 39 & 39 & 52 \\ 
                           & broad-band  & $\sim$550 & $-$145 & $-$138 & 21 & 27 & 35 & 41 \\\hline  
\rule{0ex}{3ex}175         & narrow-band & $\sim$360 &        &        &    &    &    &    \\ 
                           & narrow-band & $\sim$550 & $-$126 & $-$107 & 30 & 49 & 30 & 49 \\ \hline\\
\multicolumn{9}{l}{$^\mathrm{(1)}$ As measured with an average detector at a distance of 10~m with given detector bandwidths, $\Delta f_\mathrm{det}$.}\\
\end{tabular}
\end{table*}

The maximum detected spectral power flux densities were about 500~Jy (average over one spectral channel) for the narrow-band signals and of the order of a few Jy for the broad-band signals. As the distance to the satellites, $d$, and the main beam gain of the HBA TAB are known, it is possible to determine the transmitter spectral EIRP (equivalent isotropically radiated power), $P_\nu^\mathrm{tx}$. The EIRP is the power that a transmitter with an isotropic antenna would have to radiate to produce the observed signal. As the transmitter antenna pattern, $G_\mathrm{tx}$, and pointing direction are unknown, it is not possible to infer the conducted power at the antenna port of the satellite. The conversion formula between spectral EIRP and measured power flux densities is given by
\begin{equation}
    S_\nu = G_\mathrm{tx}(\vartheta, \varphi)\frac{P_\nu^\mathrm{tx}}{4\pi d^2} \overset{G\equiv1}{=}  \frac{P_\nu^\mathrm{tx}}{4\pi d^2}\,,
\end{equation}
assuming only line-of-sight propagation loss and neglecting other effects, such as atmospheric attenuation. The resulting minimum and maximum spectral EIRP values for each band are compiled in Tab.~\ref{tab:sat_tx_parameters}, providing results for low- and high-altitude satellites separately.

The transmitted EIRPs can also be converted to electric field strengths to make comparison with EMC standards simpler; compare Section~\ref{subsec:epfd_simulation_setup}. The corresponding values are also provided in the table. For the narrow-band signals at 125, 135, 143, 150, and 175~MHz, respectively, electric field strengths in the range of 24 to 49~$\mathrm{dB}\left[\mu\mathrm{V}\,\mathrm{m}^{-1}\right]$ are determined, normalised to what an average detector with bandwidth of 120~kHz at a distance of 10~m would measure. The typical values for the broad-band signals are between 21 and 39~$\mathrm{dB}\left[\mu\mathrm{V}\,\mathrm{m}^{-1}\right]$, again for a 120~kHz detector bandwidth. 

These values can be compared with the results of the EPFD simulations in Section~\ref{subsec:epfd_simulation_results}, in particular with Table~\ref{tab:epfd_results}. In the EPFD simulations it was however assumed that a signal had a constant electrical field strength over the full allocated RAS band 150.05$-$153~MHz.  All electrical field values have also been converted to a measurement bandwidth of 2.95~MHz which fully covers the RAS band for the convenience of comparison. They are provided in the right-most column of Tab.~\ref{tab:sat_tx_parameters}. It is noted that for narrow-band signals the values are the same for both detector bandwidths (120~kHz and 2.95~MHz), because the total integrated power is the same, while for a broad-band signal the total power increases, the more bandwidth is considered. The range of field strengths for the measurement bandwidth of 2.95~MHz is thus 24 to 49~$\mathrm{dB}\left[\mu\mathrm{V}\,\mathrm{m}^{-1}\right]$ (narrow-band) and 35 to 52~$\mathrm{dB}\left[\mu\mathrm{V}\,\mathrm{m}^{-1}\right]$ (broad-band)

For the detected Starlink satellites, Table~\ref{tab:epfd_results} cites maximum E-field values of 25.6 and $23.8~\mathrm{dB}\left[\mu\mathrm{V}\,\mathrm{m}^{-1}\right]$ given a measurement bandwidth of 2.95~MHz for the (effective) antenna diameters of 25 and 70~m, respectively. Hence, even the weak detections exceed the suggested limit, while the brightest detections are more than 20~dB above the limit.

It has to be emphasised, though, that our observations represent only a snapshot, measuring a small sub-set of all satellites and that the detected signals are not equally bright and some satellites did not even reveal UEMR at certain frequencies. Nevertheless, the overall number of detections indicates that satellite-borne UEMR from large satellite constellations could indeed be an issue for RAS operations.

\subsection{Intrinsic emission or reflection?}
Theoretically, it is possible that the measured signals do not originally stem from Starlink satellites but are of terrestrial origin, reflected off the satellites. To test this hypothesis, we first determine whether a terrestrial signal could only be visible as reflection, but not over the direct terrestrial path. Second, the transmitted power level is estimated, which would be required to create a signal of the observed properties.

\subsubsection{Geometrical considerations}
Before the link budgets of both propagation paths can be compared, the geometry of the paths needs to be worked out. The highest likelihood that a terrestrial transmitter at distance $d$ from the RAS station is not seen, while the reflected signal is visible, is given when $d$ is as large as possible compared to transmitter$-$satellite and satellite$-$receiver distance, $d_1$ and $d_2$ respectively. This is the case, when all three objects (transmitter, satellite, and receiver) are in a plane perpendicular to the ground. It is noted that none of the paths actually follow straight lines. The terrestrial path, $d$ follows a geodesic, while $d_1$ and $d_2$ are subject to refraction (which was not considered in this analysis).

\begin{figure}[!tp]
    \centering
    \includegraphics[width=\columnwidth]{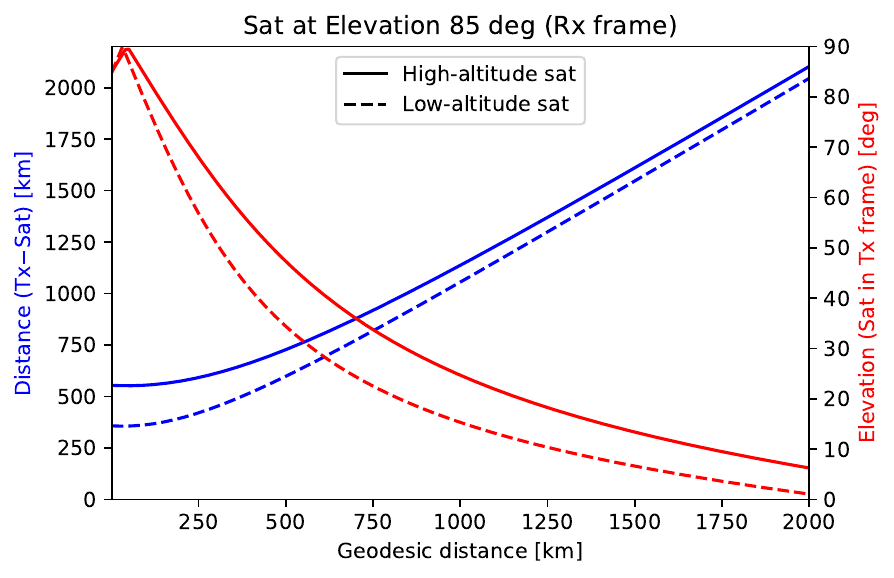}
    \caption{Geometry of the satellite reflection scenario.}
    \label{fig:radar_geometry}
\end{figure}

In Fig.~\ref{fig:radar_geometry} the path geometry is analysed for the high- and low-altitude satellites. It is assumed that the satellite appears at an elevation angle of $85\deg$ from the LOFAR observer. Based on the azimuthal angle of the satellite (with respect to LOFAR) one can construct a 
geodesic\footnote{For simplicity, the Earth Ellipsoid WGS-84 is assumed.} starting at the LOFAR observer out to a certain distance. Along this path, one can put a hypothetical transmitter and determine under which elevation angle the same satellite would appear in the transmitter frame (topocentric). Likewise, the geodesic distance (i.e. the projection on the ground) between transmitter and satellite can be inferred. The latter two quantities are shown in Fig.~\ref{fig:radar_geometry} as red and blue curves, respectively. At about 2000~km distance, the low-altitude satellite would be set below the horizon.

\subsubsection{Link budgets}
The propagation losses for both paths are determined by different physical processes. In the terrestrial case, the diffraction on the spherical Earth, tropospheric scatter, and other effects play a role. The model proposed in \citetalias{itu_p452_17} is employed to calculate the loss, $L_\mathrm{terr}(d)$. For the effective propagation loss, also the antenna gains need to be considered:
\begin{equation}
    \frac{P_\mathrm{rx}}{P_\mathrm{tx}} = G_\mathrm{tx} G_\mathrm{rx} L^{-1}_\mathrm{path}\,.\label{eq:terrestrial_pathloss}
\end{equation}
In the line-of-sight case (which is not relevant here), one would find\footnote{Because $P_\mathrm{rx} = S\cdot A_\mathrm{eff}^\mathrm{rx} = \frac{P_\mathrm{tx}G_\mathrm{tx}}{4\pi d^2} A_\mathrm{eff}^\mathrm{rx}$ and $A_\mathrm{eff}^\mathrm{rx} = G_\mathrm{rx}\frac{\lambda^2}{4\pi}$.}
\begin{equation}
L_\mathrm{terr}(d) \approx \left[\frac{4\pi d}{\lambda}\right]^2\,,
\end{equation}
It should be pointed out that we follow the common practice of spectrum management and many other fields, to define the loss as a quantity larger than One (i.e. positive on the Decibel scale).

Unfortunately, it is not known, what the antenna gains towards the local horizon are for both, transmitter and receiver. Therefore, we have to assume values. The most simple choice is to set both gains to 0~dBi.

For the reflection scenario, the Radar equation has to be used:
\begin{equation}
    P_\mathrm{rx} = \frac{P_\mathrm{tx}G_\mathrm{tx}}{4\pi d_1^2}\sigma_\mathrm{rc} \frac{1}{4\pi d_2^2}  A_\mathrm{eff}^\mathrm{rx}\,,
\end{equation}
and we can express this in a similar way as Eq.~\ref{eq:terrestrial_pathloss}:
\begin{equation}
\begin{split}
    \frac{P_\mathrm{rx}}{P_\mathrm{tx}} 
    & = G_\mathrm{tx} G_\mathrm{rx} \left[\frac{4\pi}{\lambda^2}\right] \left[\frac{\lambda}{4\pi d_1}\right]^2 \sigma_\mathrm{rc} \left[\frac{\lambda}{4\pi d_2}\right]^2 \\
    & \equiv G_\mathrm{tx} G_\mathrm{rx} \left[\frac{4\pi}{\lambda^2}\right] L^{-1}_\mathrm{sky}(d_1) \sigma_\mathrm{rc}  L^{-1}_\mathrm{sky}(d_2)\,.
\end{split}
\end{equation}
Here, the radar cross section, $\sigma_\mathrm{rc}$, was introduced. For Starlink, we assume $\sigma_\mathrm{rc}=10~\mathrm{m}^2$ as we are not aware of a publicly available measurement. The effective cross section also depends on the orientation of the satellite and the frequency range considered. Note, that for a mono-static Radar, $d_1=d_2=d$, and thus the propagation loss would scale with distance to the fourth power. In our case however, $d_1$ and $d_2$ can be very different.

\begin{figure}[!tp]
    \centering
    \includegraphics[width=\columnwidth]{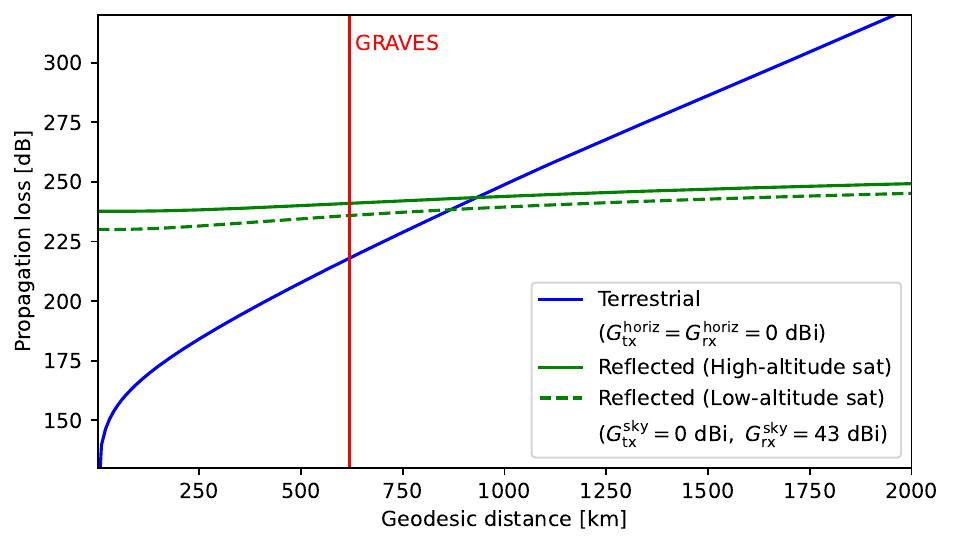}
    \caption{Path propagation losses of the satellite reflection scenario vs. the direct terrestrial (trans-horizon) path loss. The red line indicates the distance ($\sim620~\mathrm{km}$ of the GRAVES radar from the LOFAR observer.}
    \label{fig:radar_pathlosses}
\end{figure}

Figure~\ref{fig:radar_pathlosses} displays the path propagation losses of the satellite reflection scenario vs. the direct terrestrial (trans-horizon) path loss. It has to be noted that for the terrestrial path, neither the terrain (such as hills) nor clutter was accounted for. Both can add substantial additional path propagation losses of 20~dB and more, each. In the reflection case, the LOFAR TAB points at the satellite, such that the full main beam gain applies (43~dBi at 175~MHz). Again, without further knowledge it is assumed that the transmitter gain towards the satellite is 0~dBi. Under these assumptions, the propagation path over the reflection off the satellite would be more efficient beyond about 900~km compared to the terrestrial propagation. At this distance the satellite would appear at approximately 20$-$25$\degr$ elevation in the transmitter frame. If the transmitter signal would be targeted towards the satellite, then the antenna gain, $G_\mathrm{tx}^\mathrm{sky}$ would be much higher than the assumed 0~dBi, which would further decrease the distance at which the reflection scenario is more efficient. Likewise, diffraction at terrain or clutter losses would also increase the terrestrial path loss and make the reflection scenario more efficient.

\subsubsection{Estimating the transmitter power (reflection scenario)}
The calculations above show that it is indeed possible for a transmitter to create a stronger reflected signal than over the direct terrestrial path, once the distance between transmitter and receiver gets large enough. This is a consequence of the large diffraction loss on the trans-horizon terrestrial path. But still, the propagation loss via the satellite reflection is very high, so it may be interesting to estimate the required transmitter power. Again, as the transmitter antenna gain is unknown, we can only calculate the EIRP (towards the satellite), but not the conducted power at the antenna port of the transmitter.

Based on the reflected-case path propagation loss in Fig.~\ref{fig:radar_pathlosses} and the maximum received narrow-band power of 
\begin{equation}
    P_\mathrm{rx}=A_\mathrm{eff}^\mathrm{tab}S_\nu\Delta f = 2997~\mathrm{m}^2\cdot 506~\mathrm{Jy}\cdot  12.2~\mathrm{kHz}=-157~\mathrm{dB[W]}\,,
\end{equation}
the transmitter power (EIRP towards satellite) would need to be between 81~and 92~$\mathrm{dB[W]}$ or 73~and~88~$\mathrm{dB[W]}$ for high- or low-altitude satellites, respectively, depending on the distance between radar transmitter and satellite. This is a huge number and would require a LOFAR-size transmitter with a conducted power in the kilo-Watts regime (concentrated within a bandwidth of only 12.2~kHz)\footnote{Similar figures apply for the broad-band signals. While these have lower intensity, they span many MHz and already the fraction in the RAS band (150.05$-$153~MHz) leads to the same received power as the higher-intensity narrow-band signal.}. In the case of the GRAVES frequency it is indeed very likely that the detected signal is in fact originating from the GRAVES radar and reflected off the satellites. GRAVES probably has sufficient transmit power to explain the received signals. In fact, there are numerous reports by amateurs who receive GRAVES signals that were reflected by meteors with small receiving antennas. The distance between GRAVES in the eastern part of France and the LOFAR superterp is about 620~km. This is large enough to make the reflection path more efficient than the direct terrestrial path, as for LOFAR the local clutter environment will play a role and there is also relatively hilly terrain along the propagation path in France and Belgium that would increase the diffraction losses.

No other radar facility is known that operates at the detected frequencies. While for these cases, the radar scenario cannot be fully excluded, we find it unlikely. The fact that the 125~and 175~MHz signals were only observed for the higher orbit satellites is another aspect that would be hard to explain within a radar scenario. And a powerful radar that would operate broad-band between about 110~and 170~MHz would probably be well-known as it would interfere with many applications in a large area around the transmitter.

\section{Summary and conclusions}\label{sec:summary}
Using the LOFAR radio telescope, we have detected radiation between radio frequencies of 110 and 188 MHz that is correlated with satellites of the SpaceX/Starlink constellation. These frequencies are well below the assigned transmission frequencies at 10.7 to 12.7~GHz. Broad-band emission was present over the whole observed bandwidth for some satellites, while others showed strong (from 10~Jy up to $\sim$500~Jy) narrow-band signals at frequencies of 125, 135, 150, and 175~MHz. The presence of narrow-band emission differs between Starlink satellites at operational altitudes with those that were still actively raising their orbits, indicating possible differences in the operational state of the satellites, or differences between their hardware versions. We found that the flux density of the broad-band emission decreases with range, suggesting this emission is likely intrinsically generated and is detectable in 47 of the 68 Starlink satellites that were observed. However, narrow-band radio emission at 143.05~MHz can be attributed to reflections of transmissions from the French GRAVES space surveillance radar, and while we know of no other radars operating at the detected narrow-band frequencies or broad-band frequency ranges, confirmation that the observed narrow-band emission at other frequencies is intrinsic is required.

The narrow-band emission detected at 125, 150, and 175~MHz may be harmonically related, suggesting a local oscillator or clock signal operating at a frequency of 25~MHz. It is noteworthy that the narrow-band signals were only detected for satellites at the operational altitude. No such signals were seen for the satellites in orbit-raising phase, it is unclear if this effect is owing to operation or satellite version. The broad-band features are with high probability caused by other means, such as switched-mode power supplies, communication signals internal to the satellites, or some other electronic or electrical subsystem.

Follow-up observations will be able to shed further light on the origin and properties of the observed emission. Observations with the LOFAR Low-Band Antennas (LBAs; 10$-$90~MHz) would be able to confirm the presence of a 25~MHz local oscillator, while higher frequency resolution observations should allow the distinction between intrinsic or reflected emission from the Doppler shifts of the narrow-band emission. Further observations with LOFAR as well as other radio telescopes will be required to investigate the properties of the emission between different Starlink satellite versions at operational altitudes, if the emission changes when the satellites are in the Earth's shadow and the solar array is not illuminated by the Sun, and if radio emission from Starlink satellites is detectable at higher radio frequencies. Besides further observations of satellites from the Starlink constellation, it would be prudent to determine if satellites from other constellations emit UEMR. Finally, the impact of -- and possible mitigation strategies against -- the observed emission from satellites of the Starlink, or any other, constellation on the different science cases of LOFAR and other current, as well as future, radio observatories (e.g. MWA, LWA, SKA1-Low) operating at low frequencies needs to be investigated.

Any kind of UEMR is not subject to spectrum management of active radio services. In fact, from the radio astronomers perspective, UEMR is currently not well regulated for satellites and spacecraft. While there are some electromagnetic compatibility standards for spacecraft, these were made to protect the subsystems within a spacecraft from each other or its launcher system, but not to protect third party activities. The measurements presented in this paper show that there is a potential for harmful interference (as defined in the ITU-R radio regulations using the RA.769 thresholds) in radio astronomy observations caused by satellites in frequency bands far away from their allocated carrier frequencies. This potential is a function of the number of satellites and their orbital parameters, thus large satellite constellations may pose a risk. A big difference between wanted transmissions via antennas and UEMR, is that the latter is most likely not directional but relatively isotropic. Therefore, one important protection measure, which is to exclude radio astronomy stations from the service area of a satellite network, is not possible for UEMR. In addition, a strong terrestrial transmitter, which is not immediately an issue because of good geographical separation, can produce reflected signals via the satellites' surfaces. A sphere of satellites could produce a new propagation channel which may need to be considered in terrestrial radio-propagation models as the ones developed by the ITU, this requires further study. Both effects, intrinsic and reflected emission, are presently not considered in the national and international regulation processes. 

Because the detected signals in our one-hour observation represent only a snapshot and a small fraction of the Starlink constellation, one can currently not estimate accurately if and how much an entire satellite constellation, Starlink or other, would exceed protection thresholds in RAS frequency bands. However, the detected intensities are orders of magnitude above the level that each individual satellite would be allowed to have in order to comply with the \citetalias{itu_ra769_2} thresholds (if all satellites were equally bright as explained in Section \ref{sec:impact_of_emr}). Therefore, we are of the opinion that satellite operators and regulation authorities should consider satellite UEMR and reflected signals as another aspect of the regulatory process. 

Additionally, a dialogue between the satellite operators and the (radio) astronomical community would be welcome to understand how the electrical properties and operational procedures of the satellites affect radio astronomy, and how these can be used to mitigate their impact. Hopefully, this dialogue can build on the co-operation that SpaceX/Starlink has with optical astronomy \citep[see discussion in][]{green22}, especially since radio observations may be affected continuously, not primarily during twilight as is the case with optical/infrared astronomy. This could follow the example that was set with the recent coordination agreement between the US National Science Foundation (NSF) and SpaceX. Most of the authors of this work are active members in the IAU CPS, where this dialogue can take place.

It cannot be overstated, that any loss of observing time can directly be translated into a monetary loss of the substantial investments which went into developing, operating and using radio astronomy facilities \citep[e.g.][]{barentine23}. However, the much graver consequence is the loss of the output of this comparably small investment -- fundamental research is a significant sector of physical science, which usually pays out only in a matter of  decades. While some of the existing satellite constellations have the means to protect radio astronomy sites from intended radio transmissions by steering their radio beams away, this kind of active mitigation will not be possible for UEMR. Hence, this is an issue in need of close attention by satellite operators, regulators and the astronomical community. Tens of thousands of low-Earth orbit satellites are in the making and without proper consideration, these could potentially produce an artificial sphere of `radio light' that leaks into astronomical observations, rendering some astronomical observations impossible.

\begin{acknowledgements}
This paper is based (in part) on data obtained with the International LOFAR Telescope (ILT) under project code DDT16\_003. LOFAR \citep{hwg+13} is the Low Frequency Array designed and constructed by ASTRON. It has observing, data processing, and data storage facilities in several countries, that are owned by various parties (each with their own funding sources), and that are collectively operated by the ILT foundation under a joint scientific policy. The ILT resources have benefitted from the following recent major funding sources: CNRS-INSU, Observatoire de Paris and Université d'Orléans, France; BMBF, MIWF-NRW, MPG, Germany; Science Foundation Ireland (SFI), Department of Business, Enterprise and Innovation (DBEI), Ireland; NWO, The Netherlands; The Science and Technology Facilities Council, UK; Ministry of Science and Higher Education, Poland. The project leading to this publication has received funding from the European Union’s Horizon 2020 research and innovation programme under grant agreement No 101004719.

The authors thank the support of the IAU Centre for the Protection of the Dark and Quiet Sky from Satellite Constellation Interference (IAU CPS). The IAU CPS is a virtual centre of the International Astronomical Union set up in partnership with the SKAO and the NSF’s NOIRLab. The Centre coordinates collaborative and multidisciplinary international efforts from institutions and individuals working across multiple geographic areas, seeks to raise awareness, and mitigate the negative impact of satellite constellations on ground-based optical, infrared and radio astronomy observations as well as on humanity’s enjoyment of the night sky. Conversations about UEMR started back in 2020 on the Dark and Quiet Skies 2 workshop and this paper is a result of those conversations and studies.

We thank Willem Baan and Uwe Bach for proof-reading our initial draft and providing valuable feedback.

This paper made extensive use of the Python scientific stack, and we would like to thank the developers of NumPy \citep{NumPy}, matplotlib \citep{Matplotlib}, SciPy \citep{SciPy}, Astropy \citep{Astropy2013,Astropy2022}, and Cython \citep{Cython}. 

\end{acknowledgements}

\bibliographystyle{aa}
\bibliography{references}

\begin{appendix}
    
\section{The Equivalent-Power Flux Density method (EPFD)}\label{appendix:epfd}
Mathematically, the received aggregated power for a RAS pointing direction, $(\varphi_0, \vartheta_0)$, is given by
\begin{equation}
    P_\mathrm{rx}(\varphi_0, \vartheta_0)=\sum_{i=0}^{n} L_i^{-1}(\varphi_i, \vartheta_i, d_i) G_\mathrm{rx}(\varphi_i, \vartheta_i; \varphi_0, \vartheta_0)G_\mathrm{tx}(\tilde\varphi_i, \tilde\vartheta_i) P_\mathrm{tx}\,. \label{eq:prx_agg}
\end{equation}
The angles $(\varphi_i, \vartheta_i)$ describe the position of satellite $i$ in the observer frame (e.g. azimuth and elevation), while $(\tilde\varphi_i, \tilde\vartheta_i)$ is the position of the observer in the satellite antenna frame. The distance between each of the satellites and the observer is denoted as $d_i$. Furthermore, $P_\mathrm{tx}$ is the transmitted power in forward direction, $G_\mathrm{tx,rx}$ are the effective transmitter and receiver antenna gains. The path attenuation/path propagation loss is subsumed into $L_i(\varphi_i, \vartheta_i, d_i)$.\footnote{It is common to define the path propagation loss as a quantity larger than One, such that it is positive on the Decibel scale, which is why one has to divide by $L_i$ in Eq.~\ref{eq:prx_agg}.} If only line-of-sight loss would be accounted for (which is approximately correct at low frequencies), $L_i$ becomes
\begin{equation}
    L_i^{-1}(d_i) = \left[\frac{1}{4\pi}\frac{c}{d_i f}\right]^2\,.
\end{equation}
where $c$ is the speed of light and $f$ is the observing frequency. At higher frequencies, atmospheric attenuation plays an important role, too. 

The received aggregated power as given in Eq.~\ref{eq:prx_agg} is not the quantity, which is used in \citetalias{itu_s1586_1,itu_m1583_1}. Instead, in these recommendations the EPFD is defined as
\begin{equation}
    \mathrm{EPFD}(\varphi_0, \vartheta_0)=\sum_{i=0}^{n} \frac{1}{4\pi d_i^2} \frac{G_\mathrm{rx}(\varphi_i, \vartheta_i; \varphi_0, \vartheta_0)}{G_\mathrm{rx}^\mathrm{max}}G_\mathrm{tx}(\tilde\varphi_i, \tilde\vartheta_i) P_\mathrm{tx}\,. 
\end{equation}
This assumes pure line-of-sight propagation losses. In this case, we can also identify
\begin{equation}
    \mathrm{EPFD}(\varphi_0, \vartheta_0)=4\pi\frac{f^2}{c^2}\frac{1}{G_\mathrm{rx}^\mathrm{max}}P_\mathrm{rx}(\varphi_0, \vartheta_0)\,,
\end{equation}
but as mentioned above, it is usually desired to normalise this to a hypothetical isotropic receiver, to make the comparison with \citetalias{itu_ra769_2} easier, that is
\begin{equation}
    \left.\mathrm{EPFD}(\varphi_0, \vartheta_0)\right\vert_{G_\mathrm{rx}^\mathrm{max}=1}=4\pi\frac{f^2}{c^2}P_\mathrm{rx}(\varphi_0, \vartheta_0)\,.\label{eq:epfd_norm}
\end{equation}

It should be noted that \citetalias{itu_ra769_2} also contains limits for the received power, such that it would equally well be possible to directly work with Eq.~\ref{eq:prx_agg}. In the following, all PFD values are to be understood in the sense of Eq.~\ref{eq:epfd_norm}. The simulations, carried out in this work, perform EPFD calculations for a grid of sky cells as proposed in \citetalias{itu_s1586_1} and \citetalias{itu_m1583_1}. The applied scheme returns cells that have approximately the same solid angles. \citetalias{itu_s1586_1} also recommends to use a random pointing of the radio telescope antenna in a given cell for each iteration, but if the grid cells are not too large the final results usually do not show significant dependence on this. Nevertheless, as this has no impact on the computational complexity, it is usually done in this way.

Radio telescope antenna patterns are very complicated depending on the fine details of the aperture. For example, primary foci are often attached to support legs, which block part of the aperture (as the primary focus installation itself). For general purpose calculations, \citetalias{itu_ra1631_0} contains a (radially symmetric) reference antenna pattern to be used in spectrum management compatibility studies, which is based on a non-blocked circular aperture.

In the EPFD calculation the transmitter gain as well as the receiver gain have to be accounted for, both being direction dependent. While the receiver gain depends only on the angular separation between a given telescope pointing and a satellite (owing to the symmetry of the \citetalias{itu_ra1631_0} pattern), the satellite transmitter antenna pattern can be a more complicated function. Therefore, for each time step (and thus satellite position), the relative position of the RAS station in the dynamic satellite antenna frame must be inferred.

\end{appendix}
\end{document}